\documentclass[preprint,unsortedaddress,amsmath,amssymb,superscriptaddress]{revtex4}
\usepackage{graphicx}
\usepackage{bm}

\usepackage{newfloat}

\DeclareFloatingEnvironment[name={Extended Data Figure}]{suppfigure}

\let\%\relax 
\DeclareFontFamily{OT1}{pxr}{}
\DeclareFontShape{OT1}{pxr}{m}{n}{<->pxr}{}
\DeclareSymbolFont{letA}{OT1}{pxr}{m}{n}
\DeclareMathSymbol{\%}{0}{letA}{`\%}
\usepackage{lineno}

\begin{document}
%\linenumbers

%\title{Genuine phase diagram of lightly doped high-$T_c$ cuprate superconductors unveiled by observing clean CuO$_2$ planes}
\title{Unveiling phase diagram of the lightly doped high-$T_c$ cuprate superconductors with disorder removed}
%\title{Fermi surface with well-defined quasiparticles formed by less than 1$\%$ carrier doping in extremely clean CuO$_2$ planes of a high-$T_c$ superconductor}

\author{Kifu~Kurokawa}
\affiliation{ISSP, University of Tokyo, Kashiwa, Chiba 277-8581, Japan}

\author{Shunsuke~Isono}
\affiliation{Department of Applied Electronics, Tokyo University of Science, Tokyo 125-8585, Japan}

\author{Yoshimitsu~Kohama}
\affiliation{ISSP, University of Tokyo, Kashiwa, Chiba 277-8581, Japan}

\author{So~Kunisada}
\affiliation{ISSP, University of Tokyo, Kashiwa, Chiba 277-8581, Japan}

\author{Shiro~Sakai}
\affiliation{RIKEN Center for Emergent Matter Science (CEMS), Wako, Saitama 351-0198, Japan}

\author{Ryotaro~Sekine}
\affiliation{Department of Applied Electronics, Tokyo University of Science, Tokyo 125-8585, Japan}

\author{Makoto~Okubo}
\affiliation{Department of Applied Electronics, Tokyo University of Science, Tokyo 125-8585, Japan}

 \author{Matthew~D.~Watson}
\affiliation{Diamond Light Source, Harwell Campus, Didcot, OX11 0DE, United Kingdom}

\author{Timur~K.~Kim}
\affiliation{Diamond Light Source, Harwell Campus, Didcot, OX11 0DE, United Kingdom}
   
\author{Cephise~Cacho}
 \affiliation{Diamond Light Source, Harwell Campus, Didcot, OX11 0DE, United Kingdom}
 
 \author{Shik~Shin} 
\affiliation{ISSP, University of Tokyo, Kashiwa, Chiba 277-8581, Japan}
\affiliation{Office of University Professor, University of Tokyo, Kashiwa, Chiba 277-8581, Japan}

\author{Takami~Tohyama}
\affiliation{Department of Applied Physics, Tokyo University of Science, Tokyo 125-8585, Japan}

\author{Kazuyasu~Tokiwa} 
\email{tokiwa@rs.tus.ac.jp}
\affiliation{Department of Applied Electronics, Tokyo University of Science, Tokyo 125-8585, Japan}

\author{Takeshi~Kondo}
\email{kondo1215@issp.u-tokyo.ac.jp}
\affiliation{ISSP, University of Tokyo, Kashiwa, Chiba 277-8581, Japan}
\affiliation{Trans-scale Quantum Science Institute, The University of Tokyo, Bunkyo-ku, Tokyo 113-0033, Japan}
\date{\today}

\maketitle
{\textbf {Abstract}} 

The currently established electronic phase diagram of cuprates is based on a study of single- and double-layered compounds. These CuO$_2$ planes, however, are directly contacted with dopant layers, thus inevitably disordered with an inhomogeneous electronic state. Here, we solve this issue by investigating a 6-layered Ba$_2$Ca$_5$Cu$_6$O$_{12}$(F,O)$_2$ with inner CuO$_2$ layers, which are clean with the extremely low disorder, by angle-resolved photoemission spectroscopy (ARPES) and quantum oscillation measurements. 
We find a tiny Fermi pocket with a doping level less than 1~$\%$ to exhibit well-defined quasiparticle peaks which surprisingly lack the polaronic feature. This provides the first evidence that the slightest amount of carriers is enough to turn a Mott insulating state into a metallic state with long-lived quasiparticles. By tuning hole carriers, we also find an unexpected phase transition from the superconducting to metallic states at 4~$\%$. Our results are distinct from the nodal liquid state with polaronic features proposed as an anomaly of the heavily underdoped cuprates.

\newpage
{\textbf {Introduction}} 

Over 30 years of research on the cuprates has led to a ``unified” form of the phase diagram, supposed to be applicable to various cuprates \cite{Keimer2015}. According to it, the Mott insulator with the antiferromagnetic (AF) order persists up to about 5~$\%$ of carrier doping ($p\sim0.05$), followed by a dome-shaped superconducting (SC) phase; these two phases are clearly separated without the slightest overlap. In the underdoped region, the pseudogap \cite{Yasuoka,Timusk} and charge-density-wave (CDW) \cite{Wise_CDW} states competes with superconductivity \cite{Kondo,Chang,Damascelli_CDW,Hashimoto,Ghiringhelli}. These states develop most significantly around the antinode [or ($\pi,0$) region], leaving only arc-like segments of the Fermi surface (FS) even above $T_c$ \cite{Norman}. As the hole doping decreases, the Fermi arc shrinks and eventually becomes point nodes (or nodal liquid state) at the edge of the Mott insulating phase \cite{Kanigel2006,Yoshida_LSCO,Zhou_Bi2201,Shen2004,Chatterjee}. In this state, the quasiparticle peak is tiny and accompanied by polaron-like broad spectra. In addition, this peak disappears with leaving off the node since very broad spectra severely damped by the pseudogap prevail everywhere in momentum space. At doping levels further less, a gap is opened even in the (0,0)-($\pi,\pi$) direction \cite{Zhou_Bi2201,Shen2004,Vishik}, turning the nodal liquid state into a full gap state consisting only of polaronic broad spectra all around the Brillouin zone (BZ). 

Notably, the above phase diagram is based on the data of single- and double-layered cuprates;
it is different from the phase diagrams with a significant overlap of the AF and SC phases in multilayer systems with three or more CuO$_2$ planes per unit cell \cite{Mukuda,Shimizu}.
In the single- and double-layered cuprates, the CuO$_2$ plane is affected by the random potential induced by the adjacent dopant layers, leading to an inhomogeneous electronic state as revealed by scanning tunneling microscopy (STM)  \cite{STM,Pan_STM}. It is, therefore, possible that the phase diagram is relevant only for disordered CuO$_2$ planes, especially in the lightly doped region sensitive to disorder. This circumstance may have hindered a fair comparison of the data with the theory describing the doped Mott state, which usually supposes an ideal CuO$_2$ plane without disorder \cite{AFSC_RMP,AFSC_PRL}. It could, however, be solved by focusing as a research target on the inner planes of the multilayer cuprates,  which are protected by the outer CuO$_2$ planes screening the disorder effect from the dopant layers. According to nuclear magnetic resonance (NMR) studies \cite{Mukuda,Shimizu}, the carrier doping in the inner planes is much more homogeneous than that in the CuO$_2$ planes of other compounds, including the single-layered HgBa$_2$CuO$_{4+\delta}$ (Hg1201) and double-layered YBa$_2$Cu$_3$O$_{6.5}$ (Y123) thought as to be clean systems. With this advantage, the small Fermi Pocket, which had been elusive while predicted in the doped Mott state, was recently observed in the inner plane of a 5-layer compound \cite{Kunisada2020}. 
The multilayer cuprates, therefore, provide an elleclent platform to unveil the genuine electronic properties of the lightly-doped region, which is key to elucidating the pairing mechanism in cuprates. Above all, since the highest achievable $T_c$ among the existing substances is obtained in one of the multilayer cuprates (the trilayer HgBa$_2$Ca$_2$Cu$_3$O$_{8+\delta}$ \cite{pressureNature,Proust,Yamamoto}), the current subject is crucial for the development of condensed matter physics. 

%In particular, the systems with many layers have a huge advantage for studying the lightly-doped Mott state since the inner CuO$_2$ planes naturally realize an extremely clean, low doping system. 

In this article, we have selected the 6-layer Ba$_2$Ca$_5$Cu$_6$O$_{12}$(F,O)$_2$ (\textit{T}$_c$ = 69 K; Supplementary Fig. S1) for a study, where the effective carrier doping of the inner planes should be very low. The electronic properties of the clean CuO$_2$ planes are revealed over a wide range of hole doping which is controlled by varying the number of inner planes and the \textit{in situ} potassium deposition on the sample surface. The spectra with well-defined quasiparticle peaks lacking the polaronic features are detected all over the closed Fermi surface (or a tiny Fermi pocket), even at the doping level extremely close to the half-filling; it is distinct from the nodal liquid state and the polaronic state established in the heavily underdoped cuprates with the inevitable disorder. Furthermore, we find that the superconducting pairing occurs at $\sim$4~$\%$ doping, almost the same critical doping as in single-layered cuprates with CuO$_2$ planes severely disordered. This doping level ($\sim$4~$\%$), therefore, is not the consequence of an increase by the disorder but should be the critical amount of carriers essential for the pair formation even in the ideally clean CuO$_2$ plane. 

{\textbf {Results}}

Figure 1a plots the spectral intensities close to the Fermi level ($E_F$) measured by laser-ARPES at the lowest temperature (\textit{T} = 5 K). We found three sheets of FSs: One exhibits an arc-like structure typical for the underdoped cuprates, and the other two show small pockets around ($\pi/2,\pi/2$) corresponding to the doped Mott states with the AF order~\cite{Marino_2020}. To further validate the ARPES results, we also observed the de Haas-van Alphen (dHvA) effect by the torque measurement, a bulk-sensitive probe. We detected quantum oscillations (Fig. 1c) consisting of mainly two frequencies (arrows in Fig. 1f), corresponding to the FS areas covering 1.2 and 4.8~$\%$ of the Brillouin zone. These values almost perfectly agree with the ARPES results (1.0 and 4.3~$\%$). Since carriers are doped from the dopant layers (hatched by orange in Fig. 1f), the doping amount should become less toward the inner planes. It is, thus, expected that the small Fermi pocket, large Fermi pocket, and Fermi arc are each formed by the inner-most layer (IP$_0$), second-inner plane (IP$_1$), and outer plane (OP), respectively, as noted in Fig. 1a and Fig. 1f. 
The validity of this one-to-one correspondence between FSs and CuO$_2$ layers can be confirmed by comparing these results with those of the 5-layer compound \cite{Kunisada2020}, as demonstrated below.

In Fig. 1b, we overlay the FSs of the 5- and 6-layer compounds determined from the laser-ARPES data. Here, note that the samples we observed have similar \textit{T}$_c$ values (\textit{T}$_c$ = 65 K and 69 K for 5- and 6-layer compounds, respectively), and thus these two should have similar doping levels. We found that the small Fermi pocket gets much smaller than that of the 5-layer compound, while the other FSs (large Fermi pocket and Fermi arc) remain almost the same. We also obtained the results supporting this conclusion by synchrotron-ARPES (Supplementary Fig. S2) with higher photon energies more generally used in the cuprate research. As summarized in Fig. 1e, the area of the small Fermi pocket labeled as IP$_0$ is changed by adding one more CuO$_2$ plane in the unit cell from five to six, and importantly, the area of the 6-layer compound gets half that of the 5-layer compound (arrows in Fig. 1e). This indicates that the carriers in IP$_0$ are simply split into two for IP$_0$s doubled in the 6-layer compound (arrows in Fig. 1f) without affecting other planes (IP$_1$ and OP); this means that the wave function of IP$_0$ is independent of those of IP$_1$ and OP. It is further justified by our experimental result that the superconducting gap is observed on the Fermi pocket of IP$_1$ but not of IP$_0$ (Supplementary Fig. S4). 
The mixing of layers should produce superconducting gaps of similar magnitudes, so our data against it indicate that the two pockets derive each from different layers (IP$_0$ or IP$_1$) that are essentially electronically decoupled. 
%the two pockets disagreeing with it should be independently formed by IP$_0$ and IP$_1$.
%and only the former is situated outside the superconducting dome in the phase diagram.

We perform a model calculation based on our ARPES results and demonstrate that the mixing of layers is, indeed, negligible (Supplementary Fig. S3). The hybridization among layers is prevented by a potential difference induced by the carrier distribution along the $c$-axis. The inclusion of an interlayer hopping parameter (V) in the calculations doubly splits the small Fermi pocket, similarly to the bi-layer splitting observed in double-layer cuprates \cite{Feng2001}. This is attributed to double IP$_0$s adjacent to each other and sensitive to the V parameter. In contrast, splitting does not appear in the large Fermi pocket even for a relatively large value of V since double IP$_1$s are structurally separated. Notably, splitting of the small Fermi pocket (IP$_0$) is not experimentally observed both in ARPES and dHvA. We also note that the peak of the fast Fourier transformation (FFT) spectrum of the dHvA effect for the small Fermi pocket is relatively sharp in width (at least, sharper than the spectra of Y123 \cite{YBCO_quantum} and Hg1201 \cite{Hg_quantum}), and it is almost the same as that for the large Fermi pocket. It indicates that there is not even the slightest splitting in the small Fermi pocket, so the interlayer hopping should be negligibly small in our samples. 
This is a remarkable feature of a low doping state, and compatible with the observation in Bi2212 that the bilayer splitting energy gets smaller with decreasing carrier concentration \cite{Anzai2010}. Calculations with such a small V estimate the mixture of wave functions from different CuO$_2$ planes to be negligible (less than 3~$\%$; Supplementary Fig. S3d). This leads us to conclude that the three FSs we observed are independently formed by three different CuO$_2$ planes (IP$_0$, IP$_1$, and IP$_2$). We emphasize that this is a new aspect of multilayered cuprates that was clarified only through the measurement of the 6-layer compounds with double IP$_0$s. 

The above argument allows us to estimate the carrier concentration of each CuO$_2$ plane directly from the area of each FS. Our data indicate that the inner-most CuO$_2$ plane (IP$_0$) is doped by only 1~$\%$ of hole carriers, which is extremely close to the half-filling. In Fig. 2b and 2c, we plot the energy distribution curves (EDCs) around the small Fermi pocket for IP$_0$ (orange circles in Fig. 2a) and those symmetrized about \textit{E}$_F$ to eliminate the Fermi cut-off effect, respectively. Surprisingly, we find very sharp peaks in the spectra even for a state with such low carrier density. Here, note that the superconducting gap is absent most likely because the state of 1~$\%$ doping is situated outside the superconducting dome in the phase diagram. The spectral peak width is estimated to be about 7.1 meV (arrows in Fig. 2c), which is comparable to or even smaller than that (blue curve in Fig. 2c) of the optimally-doped Bi$_2$Sr$_2$CaCu$_2$O$_{8+\delta}$ (Bi2212) \cite{Kondo_Bi2212}, the cuprate material most well-studied by ARPES.
This means that quasiparticles as long-lived as those in the optimally doped state can develop with a tiny amount of carrier doping in an ideally clean CuO$_2$ plane.

Notably, the peak width is constant all around the Fermi pocket including hot spots at which the FS and the antiferromagnetic zone boundary (AFZB) cross \cite{Armitage_hotspot,Shen_hotspot}. %This indicates that the AF fluctuation with the ($\pi$,$\pi$) vector is totally suppressed. 
This is very different from the anisotropic nature widely acknowledged for the underdoped cuprates \cite{Timusk}. We also emphasize that the electronic state unveiled here is distinct from the following features illustrated for the single- and double-layered cuprates: the nodal liquid state, where only the nodal direction is metallic and the other $k_F$ points are dominated by broad spectra with the pseudogap \cite{Kanigel2006,Zhou_Bi2201,Shen2004,Yoshida_LSCO}, and the polaronic state, where quasiparticle peaks are tiny and largely buried by a hump-shaped incoherent part. In stark contrast, our data for inner planes exhibit surprisingly simple metallic features, forming a closed Fermi pocket (instead of the nodal liquid state) with well-defined quasiparticles (instead of the polaronic state). Here, note that the quasiparticle in IP$_0$ is not a product of the superconducting proximity effect from OP. This is evidenced by the fact that quantum oscillations (signals of well-defined quasiparticles) have been observed under conditions that the superconductivity is completely suppressed. 
ARPES also confirmed that the quasiparticle peak persists even above $T_c$ although its width gets broadened due to the thermal broadening effect (Supplementary Fig. S10). 

The bands of Fermi arc and pockets are mutually close in momentum space, so their ARPES signals could interfere at high binding energies. 
To examine single-particle spectra of IP$_0$, we conducted the band-selective measurements by utilizing the matrix element effect. In Fig. 2e, we map the FS by synchrotron-ARPES not only of the 1st BZ but also up to the 2nd BZ. The data in the 2nd BZ are enhanced in intensity, and moreover, separate the bands of Fermi arc (OP) and pockets (IP$_0$ and IP$_1$) in the upper and lower half of the 2nd BZ, respectively. This separation is further confirmed in Figs. 2f and 2g by plotting the ARPES dispersions each across nodal $k_F$ points of OP and IP$_0$ (dashed lines in Fig. 2e). Only the latter exhibits the folded bands about AFZB, which are also revealed by MDCs at $E_F$ (the upper panels of Figs. 2f and 2g). We extract the EDC for IP$_0$ at $k_F$ in Fig. 2h, and find a sharp peak accompanied by a tail with relatively low intensities up to the energy scale of the band width. This validates that the Fermi pocket possesses well-defined quasiparticles lacking polaronic features, even though the doping level is extremely small. 

In order to explore a wider doping range, we have performed the \textit{in situ} potassium deposition on the samples. This technique is commonly used in ARPES \cite{Ohta2006} including a study of cuprates \cite{Hossain2008,Damascelli_YBCO,Zhang2016}. Figures 3a to 3c display the FS mapping before and after the potassium deposition for different times (0, 30, and 60 seconds, respectively). The Fermi pockets get smaller with deposition time. This variation is more clearly demonstrated in Fig. 3f by extracting the momentum distribution curves (MDCs) along the AFZB at \textit{E}$_F$.  The momentum distance between each paired $k_F$s becomes shorter in the two pockets, representing the reduction of hole carriers in both the inner planes. In Fig. 3d and 3e, we plot the large and small Fermi pockets for IP$_1$ and IP$_0$, respectively, at three different deposition times, determined from the ARPES spectra. A systematic shrinkage of the pockets is confirmed, as estimated in Fig. 3g from their areas (or carrier concentrations $p$s). The $p$ value decreases faster in the large pocket (IP$_1$) than in the small pocket (IP$_0$); this is expected since IP$_1$ lies closer to the surface dopant layer where potassium is deposited, so it should be more efficiently doped. Our experiments could reduce the doping level of the innermost plane (IP$_0$) down to 0.7~$\%$ ($p=0.007$), which is so small as to nearly reach the half-filled Mott state.

Here we examine the doping evolution of the ARPES spectra. Figure 4a plots the EDCs at $k_F$ on the AFZB for IP$_0$ (orange circle in the inset of Fig. 4a) in 1.0~$\%$, 0.9~$\%$, and 0.7~$\%$ of doping levels. In the right panel, the symmetrized EDCs are also plotted. We found that sharp quasiparticle peaks persist down to the lowest carrier concentration of 0.7~$\%$ ($p=0.007$). Although the peak intensity was slightly suppressed due to the sample surface deterioration, the spectral peaks have almost the same width (Supplementary Fig. S8), indicating that the scattering rate (or lifetime) of quasiparticles is hardly changed with going toward the half-filling. The results imply that a single hole doped into the Mott insulator can behave as a long-lived quasiparticle in a clean CuO$_2$ plane (Fig 4b). It is in stark contrast to the property of the single- and double-layered cuprates, in which a quasiparticle peak rapidly dies out at doping levels lower than $\sim$10~$\%$ \cite{Damascelli_YBCO,Zhou_Bi2201,Shen2004,Vishik}, and even if there is some spectral weight at $E_F$, it is rather broad and disappears immediately with going away of the nodal direction \cite{Yoshida_LSCO,Devereaux}. 

IP$_1$ also realizes a very clean electronic system as confirmed by its low Dingle temperature (\textit{T}$_D$, proportional to the scattering rate) obtained by the dHvA quantum oscillations: \textit{T}$_D$ of IP$_1$ is estimated to be 12.3 K (Supplementary Fig. S5), which is between those of Y123 (\textit{T}$_D$ = 6.2 K) \cite{YBCO_quantum,Chan_NC} and Hg1201 (\textit{T}$_D$ = 18 K) \cite{Hg_quantum,Chan_NC}. Note that the carrier concentration of IP$_1$ directly estimated from the Fermi pocket area is small, only to be 4.3~$\%$, which is less than half those ($\sim$10~$\%$) of Y123 and Hg1201 samples used for the quantum oscillation measurements. Hence, the \textit{T}$_D$ of IP$_1$ is rather small, considering that it is obtained in such a low carrier concentration with a poor screening effect.

The spectra of IP$_1$ before the potassium deposition exhibit the superconducting gap consistent with the $d$-wave symmetry, being the largest at the tip of the Fermi pocket (see Supplemental Fig. S4 for more details). This data indicates the coexistence of superconductivity and AF magnetic order, similarly to the observations in the 5-layer compound by NMR~\cite{Mukuda,Shimizu} and ARPES~\cite{Kunisada2020}. Figure 4f plots the spectra at the tip of the Fermi pocket for three different doping levels (4.3~$\%$, 4.0~$\%$, and 3.6~$\%$) controlled by the potassium deposition. The superconducting gap observed at the original doping level of 4.3~$\%$ closes at 4.0~$\%$ (This is further justified in Supplementary Note 9). This indicates that the electronic state in IP$_1$ has got out of the superconducting dome and entered the metallic phase, i.e. the same as that of IP$_0$, by decreasing the hole doping.  Also, the relatively small superconducting gap observed for the pristine surface (magnitude of $\Delta_{\rm tip}$ = 5 meV at the tip of the Fermi pocket and the order parameter $\Delta_0 = 11$ meV determined by extrapolating it to the antinode) is attributed to the carrier doping level (4.3~$\%$) located almost at the edge of the superconducting dome. 
The superconducting to metal transition we observed occurs by closing a gap along the entire Fermi surface (or pocket) while maintaining well-defined quasiparticles all over it. This is very different from the phase transition for the underdoped cuprates with disordered CuO$_2$ planes, where the pseudogap accompanied by the damped broad spectra plays a significant role. Rather, our observation is quite similar to the doping-induced phase transition across the superconducting dome on the overdoped side with no indication of the pseudogap, although the Fermi surface size is apparently different, corresponding to $p$, not to 1+$p$.

In Fig .4g, we illustrate a phase diagram of lightly-doped cuprates with extremely clean CuO$_2$ planes unveiled via the direct observation of the electronic structure by ARPES. The Mott insulating state is realized only when the CuO$_2$ plane is non-doped, and is strictly half-filled; only the slightest amount of hole doping changes it to a metallic state forming a Fermi pocket with well-defined quasiparticles on the top of the lower Hubbard band (or charge transfer band). We find that the effective mass in the innermost layer (IP$_0$) and the 2nd inner layer (IP$_1$) with different carrier concentrations ($\sim$1~$\%$ and $\sim$4~$\%$, respectively) are almost the same ($\sim$0.6 $m_0$) by quantum oscillation measurements (Supplementary Fig. S7). This indicates a lack of a pronounced band narrowing when varying $p$ toward the half-filled Mott state. The transition to a Mott insulator most likely occurs by completely removing hole carriers from the lower Hubbard band until the perfect half-filling, rather than by controlling the band width. At 4~$\%$, the metal-to-superconductor transition occurs by opening the superconducting gap, and the system enters the phase where the AF order and superconductivity coexist. Intriguingly, this critical doping level is almost the same as that of some single- and double-layer compounds, such as La$_{2-x}$Sr$_x$CuO$_4$ which is known to be severely disordered according to the NMR studies \cite{Takagi1992,Yamamoto2001,Groen1990,Liang2006}. This indicates that 4~$\%$ is the intrinsic critical doping level necessary to form the superconducting pairs even in the CuO$_2$ planes with the clean-limit condition. This scenario differs from theories that suggest that superconductivity occurs simultaneously with the appearance of metallicities by carrier doping to the ideal Mott insulator without disorder \cite{AFSC_RMP,AFSC_PRL}.  Our results will provide crucial insight into understanding the intrinsic relationship between the Mott physics and the pairing mechanism in the lightly-doped CuO$_2$ planes, which has not been accessible for the single- and double-layered compounds mainly studied in the long history of cuprate research.
 
 \newpage

{\bf Methods}

{\bf Samples:} Single crystals of underdoped Ba$_2$Ca$_5$Cu$_6$O$_{12}$(F,O)$_2$ (see crystal structure in Fig. 1f) with $T_c$=69 K were grown at between 1100 $^\circ$C and 1200 $^\circ$C under a pressure of 4.5 GPa without an intentional flux. The starting composition for the crystal synthesis is Ba$_2$Ca$_3$Cu$_4$O$_{7.9}$F$_{2.1}$, which is known to be almost the same in the single crystals \cite{Tokiwa}. We have conducted X-ray diffraction measurements along the $c$-axis for all the sample pieces and confirmed that they are single crystals, not the mixtures of crystal domains with different numbers of CuO$_2$ layers per unit cell. Magnetic susceptibilities for these crystals (Fig. S1) show a sharp superconducting transition with $\sim$4 K in width, indicative of high quality in our samples; signal-to-noise ratio is not so high owing to the small volume in our crystals ($\sim 300 \times 300 \times 50$ $\mu$m in crystal size). Laue image of the single crystal (Fig. S1b) shows a four-fold rotational symmetry with no indication of structural modulations.\\ 

{\bf ARPES measurements:} 
Laser-based ARPES data were accumulated using a laboratory-based system consisting of a Scienta R4000 electron analyzer and a 6.994 eV laser (the 6th harmonic of Nd:YVO$_4$ quasi-continuous wave). 
The data presented are measured at 5 K. The overall energy resolution in the ARPES experiment was set to 1.4 meV. Synchrotron-based ARPES measurements were performed at high-resolution branch (HR-ARPES) of the beamline I05 in the Diamond Light Source, equipped with a MBS A-1 analyzer. The data presented are measured at the photon energy of 55 eV and at the temperature of 10 K. The overall energy resolution was set to $\sim$10 meV in our experiments. In both the laser- and synchrotron-ARPES measurements, a typical cleavage method was used to get a clean surface of the samples: a top post glued on the crystal is hit {\it in situ} to obtain a flat surface suitable for the ARPES measurements. The cleavage plane has been confirmed by STM to be along the F/O dopant layers \cite{Tokiwa}. \\

{\bf  Quantum oscillation measurements:}
Torque magnetometry experiments were performed with a commercial piezoresistive cantilever (SEIKO PRC-120) \cite{Torque} in pulsed magnetic fields up to 60 T (36 ms pulse duration). The cantilever directly detects the magnetic torque ($\tau$) as the result of the anisotropic magnetization of the sample, $\tau=${\boldmath $M$}$\times${\boldmath $H$}, and the magnetic quantum oscillation known as the de Haas-van Alphen (dHvA) oscillation was observed.  
Figure 1c in the main paper shows the data after subtracting background, which was obtained by fitting a quadratic function to each curve of the raw data in the range of magnetic field between 26 and 60 T.

%{\bf Data availability}\\
%The data that support the findings of this study are available from the corresponding authors upon reasonable request.\\

%\bibliography{6_layer_TK}
\newpage
%{\textbf {References}} \\

{\bf  Acknowledgements}\\
We thank M. Imada and Y. Yamaji for useful discussions, T. Yajima for technical assistance in the x-ray measurements performed using facilities of the Institute for Solid State Physics, University of Tokyo. 
We thank Diamond Light Source for access to beamline I05 under proposals SI30646, SI28930, and SI25416 that contributed to the results presented here. 
This work was supported by the JSPS KAKENHI (Grants Numbers. JP21H04439 and JP19H00651), by the Asahi Glass Foundation, by MEXT Q-LEAP (Grant No. JPMXS0118068681), and by MEXT as “Program for Promoting Researches on the Supercomputer Fugaku” (Basic Science for Emergence and Functionality in Quantum Matter Innovative Strongly-Correlated Electron Science by Integration of “Fugaku” and Frontier Experiments, JPMXP1020200104) (Project ID: hp200132/hp210163/hp220166).

\begin{figure}[htbp]
\begin{center}
\includegraphics [clip,width=15cm]{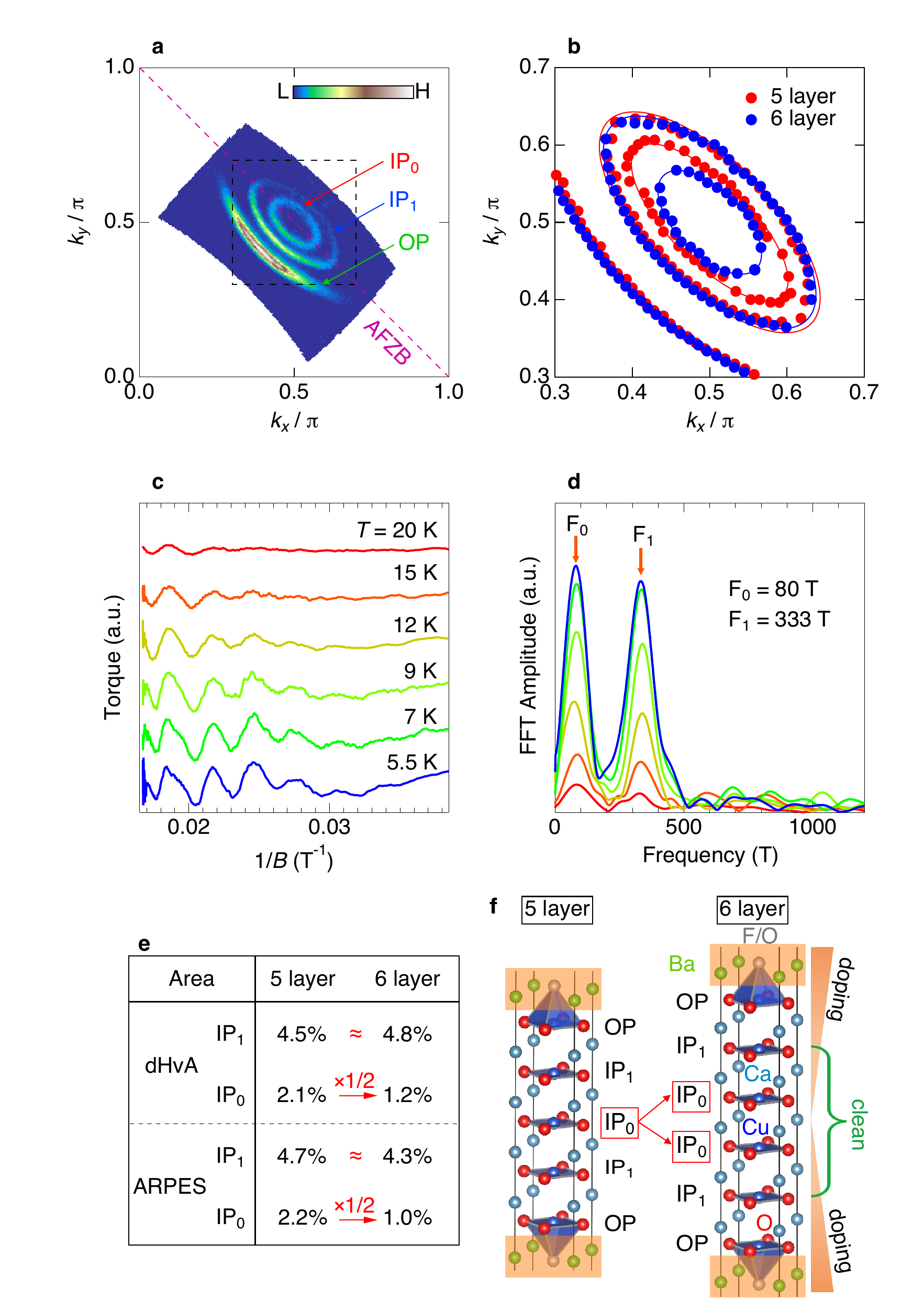}
\label{figure1}
\end{center}
\end{figure}
\clearpage

\begin{figure}
\caption{\textbf{Fermi surfaces of 6-layer cuprate and the comparison with those of 5-layer cuprate.} 
\textbf{a,} Fermi surface mapping obtained by integrating  the intensities of ARPES spectra at 5K around the Fermi energy ($E_F$). 
\textbf{b,} Fermi surfaces zoomed inside the black dot square in \textbf{a} determined from the peak positions of MDCs at $E_F$ for 6-layer (red) and 5-layer (blue) \cite{Kunisada2020} compounds. 
\textbf{c,} Quantum oscillations of the dHvA effect observed in magnetic torque signals at several temperatures. In the data, smooth background is subtracted.  The crystallographic $c$-axis was set to 2$^{\circ}$ from the magnetic field direction during the measurements.
 \textbf{d,} Fast Fourier transform spectra of \textbf{c}. The arrows show the two main peaks (F$_0$ and F$_1$), which corresponds to the two Fermi pockets observed by ARPES in \textbf{a}. 
 \textbf{e,} Comparison of the Fermi pocket areas between the 5-layer and 6-layer compounds determined by dHvA (top) and ARPES (bottom). The listed values are the area in percentage ($\%$) covering the Brillouin zone for the small and large Fermi pockets labeled as IP$_0$ and IP$_1$, respectively.  As noted with arrows, the area of the small Fermi pocket (IP$_0$) decreases to almost half with increasing the number of layers from five to six, while that of the large Fermi pocket (IP$_1$) is almost the same between the two.  
 \textbf{f,} The crystal structure of the 5-layer and 6-layer compounds. The number of the innermost CuO$_2$ plane (IP$_0$) gets doubled in the 6-layer compound, as represented by arrows.}
\end{figure}

\newpage

\begin{figure}[htbp]
\begin{center}
\includegraphics [clip,width=15cm]{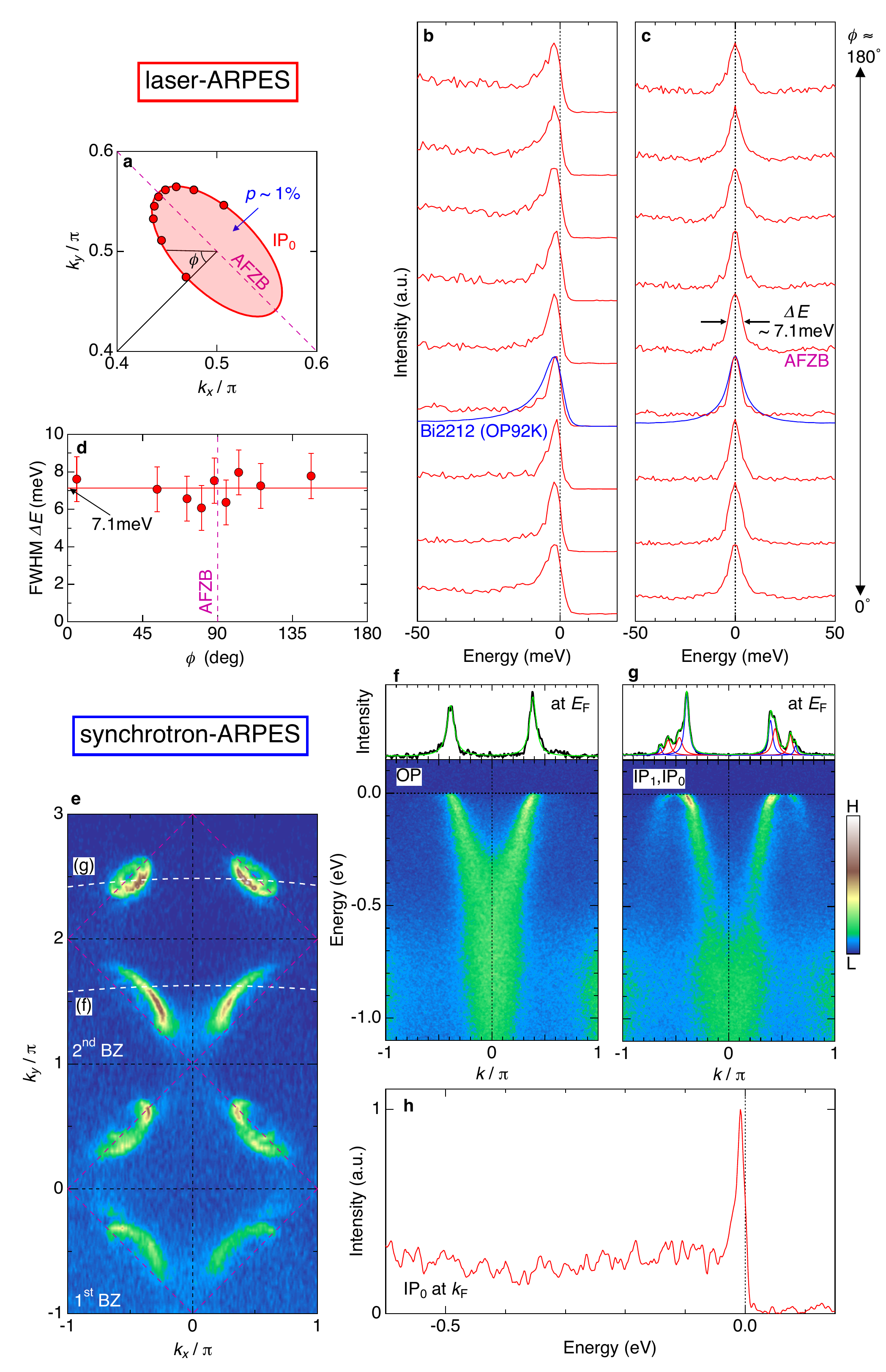}
\label{figure2}
\end{center}
\end{figure}
\clearpage

\begin{figure}
\caption{\textbf{Well-defined quasiparticle peaks without polaronic features all around the closed Fermi surface even in a slightest amount of carrier doping ($p\sim1$~$\%$).} 
\textbf{a,} The small Fermi pocket for IP$_0$ zoomed around ($\pi,\pi$). 
\textbf{b,} EDCs measured along the small Fermi pocket. The corresponding $k_F$ points are plotted by orange circles in \textbf{a}. For a fair comparison of the peak shapes, the EDCs are normalized to each peak intensity. The spectrum at the nodal point for the optimally doped Bi2212 ($T_c$ = 92 K) is overlayed (blue curve) to demonstrate that spectra of IP$_0$ in the 6-layer compound are even sharper than it.
\textbf{c,} Same data as \textbf{b}, but symmetrized about the Fermi energy to eliminate the Fermi cut-off and clarify that there is no energy gap at $E_F$ along the entire Fermi pocket. 
\textbf{d,} Angle $\phi$ (defined in \textbf{b}) dependence of the full width at half maximum (FWHM) of the symmetrized EDCs in \textbf{c} obtained by fitting to the Lorentz function.
The dashed line is the guide to the eye to represent that the spectral width is constant with the value of 7.1 meV along the Fermi pocket. Error bars represent standard deviations of the spectral peak widths.
\textbf{e,} The Fermi surface mapping up to the 2nd Brillouin zone, disentangling the Fermi arc (lower region of 2nd BZ) and pockets (upper region of 2nd BZ), by employing the matrix element effect in ARPES. 
\textbf{f,g,} The band dispersions of the Fermi arc and pockets each crossing nodal $k_F$ points of Fermi arc (OP) and small Fermi pocket (IP$_0$) 
along the white dotted lines in \textbf{e}. In upper panels, MCDs at $E_F$ are extracted to confirm that OP and IP$_0$ are indeed separately observed. The Lorentzian functions (colored curves) are fit to the data (black curve). 
\textbf{h,} The EDC at $k_F$ for IP$_0$, indicating a well-defined quasiparticle peak accompanied by the spectral tail with relatively low intensities lacking polaronic features. }
\end{figure}
\newpage

\begin{figure}[htbp]
\begin{center}
\includegraphics [clip,width=15cm]{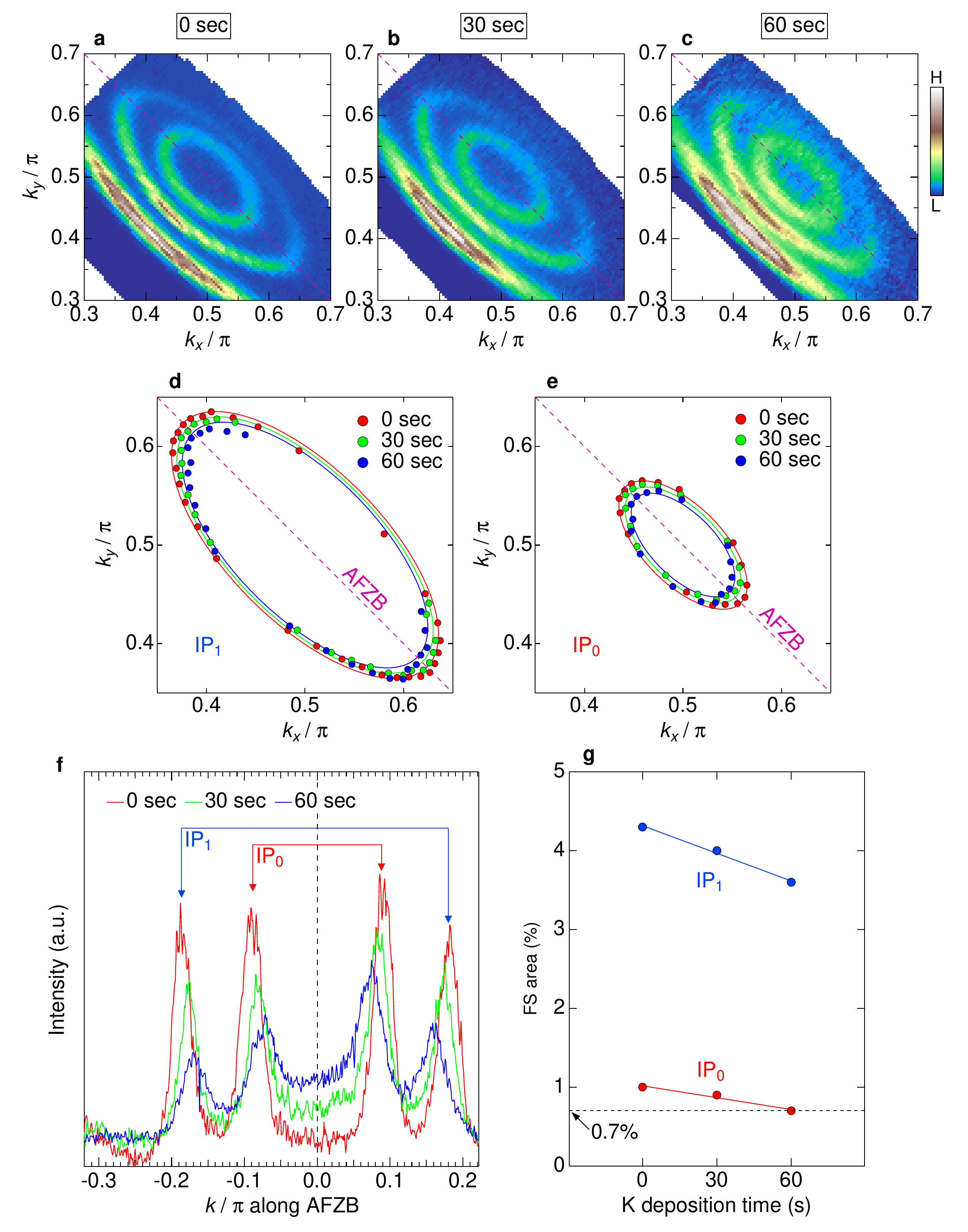}
\label{figure3}
\end{center}
\end{figure}
\clearpage

\begin{figure}
\caption{\textbf{Evolution of the Fermi pockets with \textit{in situ} potassium deposition.} \textbf{a-c,}  Fermi surface mapping zoomed around ($\pi,\pi$) to focus on the two Fermi pockets for three cases: before deposition (\textbf{a}) and after deposition for 30 seconds  (\textbf{b}) and 1 minute (\textbf{c}). The spectral intensities were integrated within an energy window of 10 meV around $E_F$. 
\textbf{d, e,} Fermi pockets determined from the peak positions of MDCs 
 for IP$_1$ (\textbf{d}) and IP$_0$ (\textbf{e}), respectively. In each panel, the results of three different deposition times (0 sec, 30 sec, and 1min) are plotted. 
\textbf{f,} Comparison of MDCs at $E_F$ along the antiferromagnetic zone boundary (AFZB). Along this momentum cut, the Fermi velocity is fastest, thus the largest shift of \textit{k}$_F$ with carrier doping is expected. 
\textbf{g,} The deposition time dependence of the area of the Fermi pocket or the hole carrier concentration $p$. The dotted line remarks that IP$_0$ reached the doping level as extremely low as 0.7~$\%$ ($p=0.007$) after the deposition for 1 minute. 
}
\end{figure}
\clearpage

\begin{figure}[htbp]
\begin{center}
\includegraphics [clip,width=16cm]{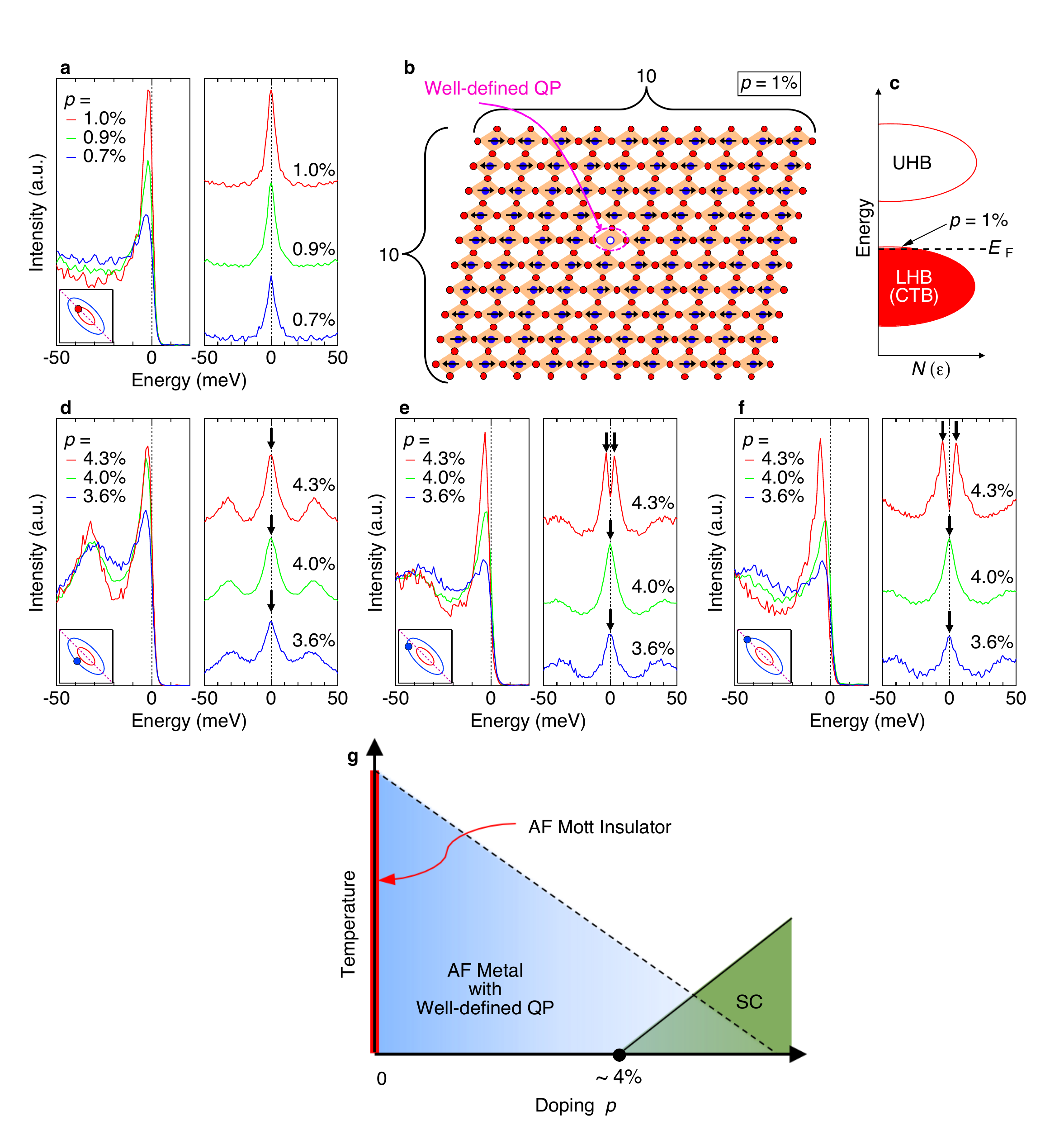}
\label{figure4}
\end{center}
\end{figure}
\clearpage

\begin{figure}
\caption{\textbf{Doping evolution of ARPES spectra and phase diagram of the lightly-doped cuprates with disorder removed.} 
\textbf{a,} Doping evolution of the energy distribution curves (EDCs) and those symmetrized about $E_F$, measured at $k_F$ on the antiferromagnetic zone boundary (AFZB) for the small Fermi pocket (orange circle at the inset). 
The spectra plotted are for three different doping levels (1~$\%$, 0.9~$\%$, and 0.7~$\%$) controlled by potassium deposition. 
\textbf{b,} Schematic of electronic state in real space for the 1~$\%$ doped CuO$_2$ plane, illustrated based on our data suggesting that a single hole can behave as a long-lived quasiparticle in the antiferromagnetic background.
\textbf{c,} Schematic density of states for the 1~$\%$ doped CuO$_2$ plane corresponding to \textbf{b}. 
\textbf{d-f,} Doping evolution of EDCs and those symmetrized about $E_F$ for three different \textit{k}$_F$s (purple circles in the inset of each panel) 
of the large Fermi pocket. The spectra plotted are for three different doping levels (4.3~$\%$, 4.0~$\%$, and 3.6~$\%$) controlled by potassium deposition. The arrows point to the spectral peaks, clarifying the gap opening or closing for each spectrum. 
\textbf{g,} The phase diagram in lightly doped region summarized from our data of clean inner CuO$_2$ planes in 6-layered cuprates. AF, QP, and SC abbreviate ``antiferromagnetic", ``quasiparticle", and ``superconductivity", respectively.}
\end{figure}
\clearpage

\end{document}

\title{Supplementary information:\\
Unveiling phase diagram of the lightly doped high-$T_c$ cuprate superconductors with disorder removed}

%\date{\today}

\maketitle

\clearpage

\begin{figure*}[h]
\includegraphics[clip,width=15cm]{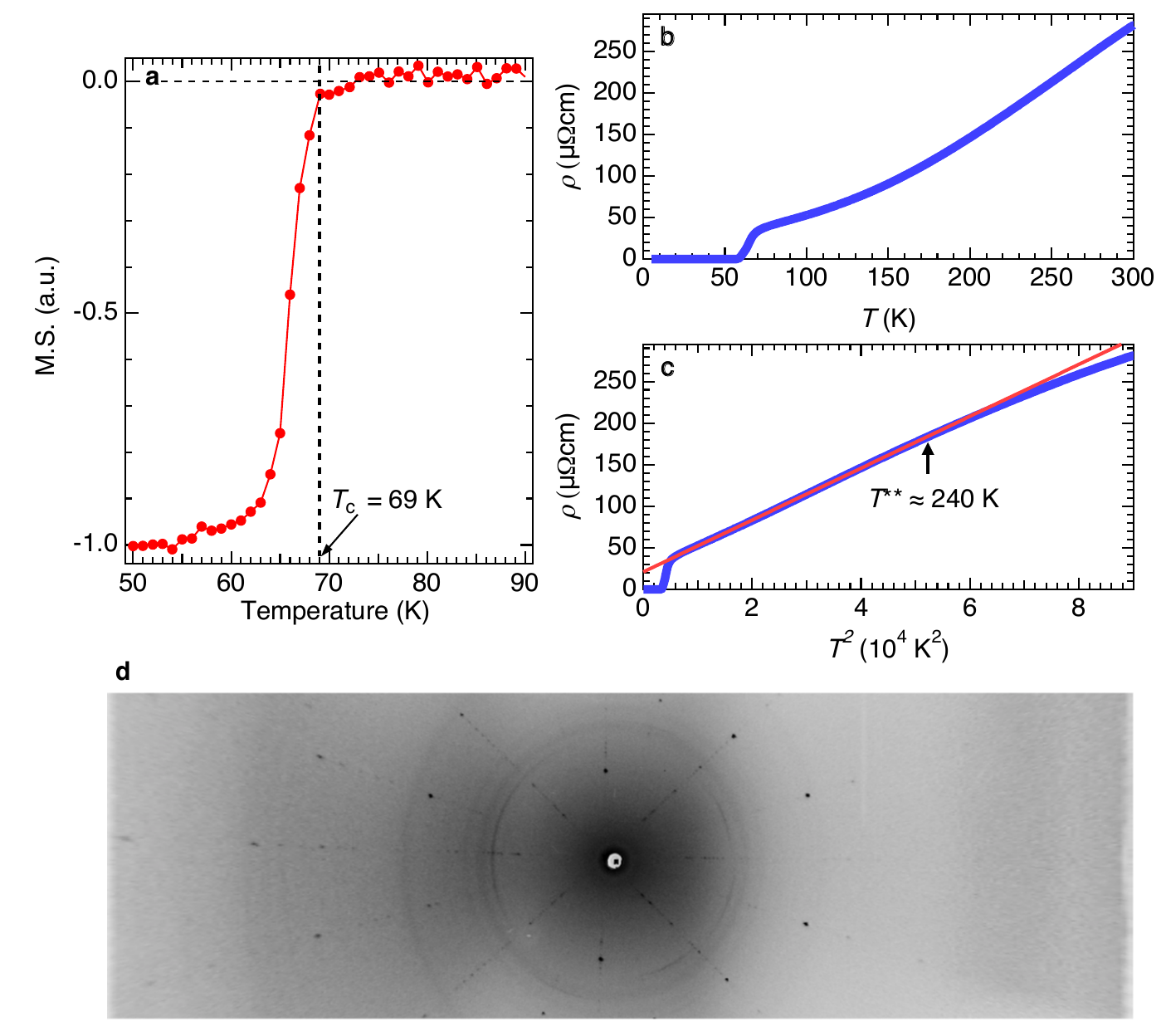}
\renewcommand{\baselinestretch}{1}
\caption{\textbf{Characterization of the 6-layer compound samples used for ARPES and quantum oscillation measurements.} 
\textbf{a.} Magnetic susceptibility. The transition temperature of 69K is estimated from the onset of Meissner effect indicated by arrow. The sharp transition with $\sim4$ K in width indicates a high quality of our samples. \textbf{b,c,} The temperature dependence of the in-plane resistivity ($\rho$) and the same data plotted as a function of $T^{2}$, respectively. The red line is fit to the $\rho$ vs. $T^{2}$ plot at low temperatures. The temperature at which the data deviate from the line is marked by an arrow at $T^{**}$ $\approx$~240 K.  
\textbf{d,} A Laue picture of the crystal, displaying a clear four-fold rotational symmetry without modulations. }
\label{Meissner}
\end{figure*}

{\noindent \fontsize{13pt}{13pt}\selectfont \textbf {Note 1: Characterization of the samples used for ARPES and quantum oscillation measurements (Fig. S1)}}
\vskip.5\baselineskip
%\section {Note 1: C\MakeLowercase{haracterization of the samples used for} ARPES \MakeLowercase{and quantum oscillation measurements} (F\MakeLowercase{ig.} \ref{Meissner})}
Figure~\ref{Meissner}a shows the magnetic susceptibility of the Ba$_2$Ca$_5$Cu$_6$O$_{12}$(F,O)$_2$ crystals we used for the current study. The transition temperature ($T_c$) is estimated to be 69 K, as the onset temperature of the Meissner effect (arrow in Fig.~\ref{Meissner}a). The transition is sharp with a width of $\sim$4 K, indicating that our samples have a high quality. We note that the signal-to-noise ratio in the magnetic susceptibility data is not so high because of the small volume in our crystals: $\sim 300 \times 300 \times 50$ $\mu$m. The in-plane resistivity (Fig~\ref{Meissner}b) shows a temperature dependence proportional to $T^2$ up to a high temperature, as demonstrated in Fig~\ref{Meissner}c ($T^{**}$$\approx$240 K marked by an arrow). This is typical behavior in the heavily underdoped regime of cuprates \cite{Greven,Ando}. Laue image of our samples (Fig.~\ref{Meissner}d) shows a four-fold rotational symmetry with no indication of structural modulations, which complicates ARPES data in momentum space. The ring-shaped background intensities come from the metal substrates on which the samples were glued.\\

{\noindent \fontsize{13pt}{13pt}\selectfont \textbf {Note 2: Observation of Fermi pockets by synchrotron-ARPES and their comparison between 5- and 6-layer compounds (Fig. S2)
}}
\vskip.5\baselineskip
%\section {Note 2: O\MakeLowercase{bservation of} F\MakeLowercase{ermi pockets by synchrotron}-ARPES \MakeLowercase{and their comparison between 5- and 6-layer compounds} (F\MakeLowercase{ig.} \ref{ARPES})}
The ARPES using a low-energy laser (laser-ARPES) has huge advantages in terms of high energy and momentum resolutions over that using synchrotron as a photon source (synchrotron-ARPES).  However, there is also a drawback in laser-ARPES: the observable momentum space is limited, not being able to cover the whole Brillouin zone (BZ), owing to the low kinetic energies of excited photoelectrons. To demonstrate the Fermi surfaces over the whole BZ, therefore, synchrotron-ARPES with higher energy photons is required, although the resolutions are sacrificed. Furthermore, synchrotron-ARPES has been more commonly used in the history of cuprate research, so confirming the consistency of its results with those by laser-ARPES presented in the main paper would be important to validate our conclusion. 

\begin{figure*}[h]
\includegraphics[clip,width=13.5cm]{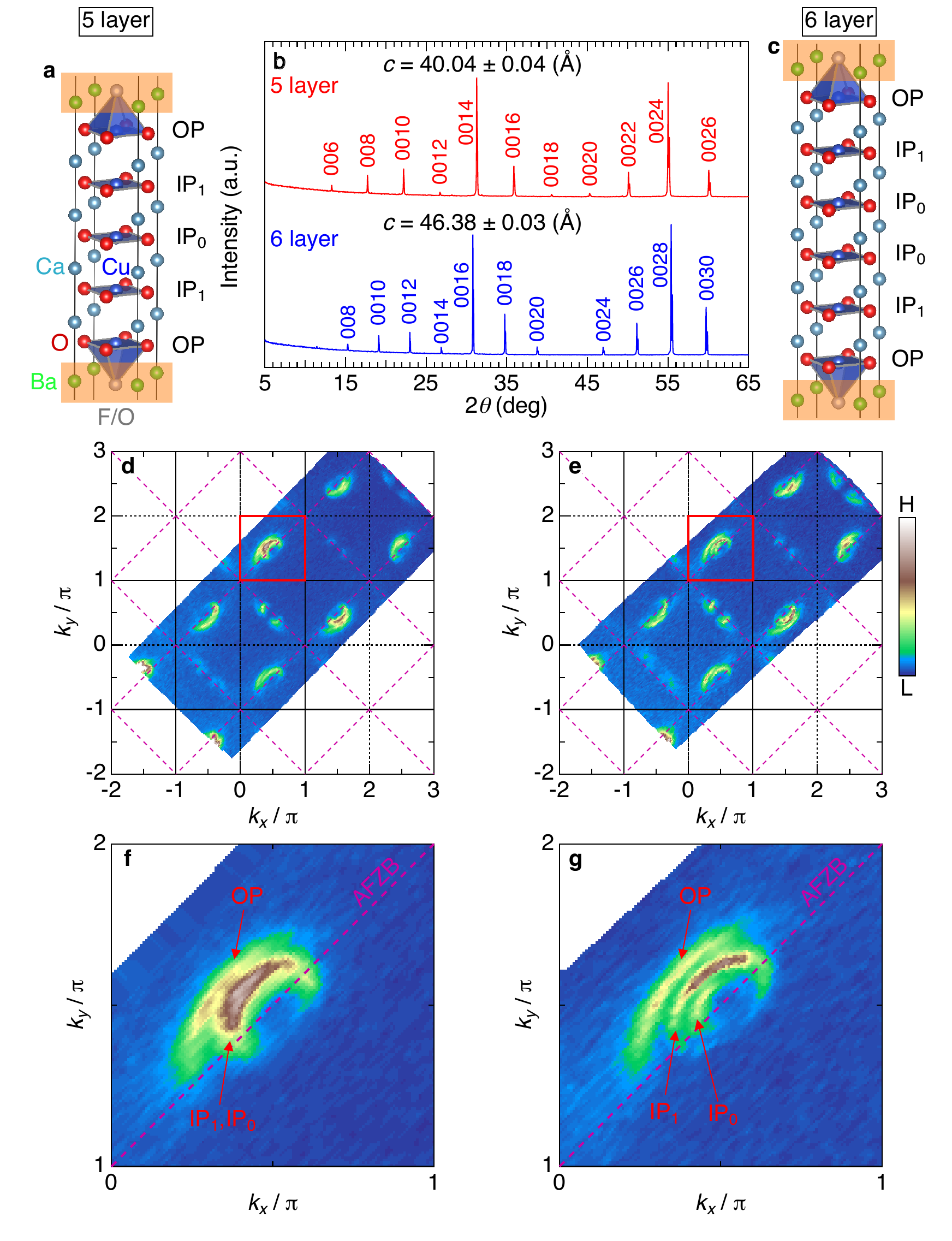}
\renewcommand{\baselinestretch}{1}
%\label{ARPES}
%\end{figure*}
%\begin{figure*}[h]
%\includegraphics[clip,width=13cm]{supplement_DLS}
\caption{\textbf{Fermi surfaces observed by synchrotron-ARPES and comparison between 5- and 6-layer compounds.} 
\textbf{a,c,} Crystal structures of 5-layer and 6-layer compounds. \textbf{b,} X-ray diffraction spectra measured along the $c$-axis, verifying the samples to be in the single phases. 
\textbf{d,e,}  Fermi surface mappings of the 5-layer and 6-layer compounds, respectively, measured at 10 K by synchrotron-ARPES with 55 eV photons. 
ARPES intensities are integrated within the energy window of 10 meV about the Fermi energy. 
\textbf{f,g,} The maps zoomed within the red squares in \textbf{d} and \textbf{e}, respectively. 
The red arrows point to the Fermi arc for the outer CuO$_2$ plane (OP) and the Fermi pockets for the inner CuO$_2$ planes (IP$_0$ and IP$_1$). 
The two pockets are clearly separated from each other in the 6-layer compound, as indicated by two arrows, whereas the ones for the 5-layer compound are mutually much closer and their spectra are overlapped. 
}
\label{ARPES}
\end{figure*}

We have prepared the underdoped single crystals of 5- and 6-layer compounds Ba$_2$Ca$_{n-1}$Cu$_n$O$_{2n}$(F,O)$_2$ ($n=5$ and 6: see each crystal structure in Figs.~\ref{ARPES}a and \ref{ARPES}c) with similar $T_c$  values of 65 K and 69 K, respectively. The X-ray diffraction spectra along the $c$-axis  (Figs.~\ref{ARPES}b) for these crystals both demonstrate commensurate layered structures, validating our crystals to be in single phases. The lattice constants are estimated as $c$~=~40.04~$\pm$~0.04~[$\AA$] and $c$~=~46.38~$\pm$~0.03~[$\AA$] for the 5- and 6-layer compounds, respectively. These values indicating the increase of about 6.37 $\AA$~in the latter due to the extra one CuO$_2$ layer are consistent with the relation of $c$~=~14.71~+~6.37$(n-1)$~[$\AA$] between $c_{0}$ and $n$ determined by the polycrystalline powder X-ray diffraction measurements \cite{Iyo2001,Iyo2006}. 

For these samples, we conducted synchrotron-ARPES measurements at $h\nu$=55 eV, and obtained 
 the Fermi surface mappings over a wide range of momentum space covering multiple BZs, as plotted in Figs.~\ref{ARPES}d and ~\ref{ARPES}e.  In both cases, Fermi pockets for the inner CuO$_2$ planes (IP$_0$, IP$_1$) and Fermi arcs for the outer CuO$_2$ planes (OP) are clearly observed not only in the 1st BZ but also in the 2nd and 3rd BZs. Due to the matrix element effect in photoemission, the intensities get weak, having structures not clear, in some regions such as around ($-\pi/2,-\pi/2$).  

For a more detailed comparison between the Fermi surfaces of 5- and 6-layered compounds, we zoom the regions marked by green squares in Figs.~\ref{ARPES}d and \ref{ARPES}e and plot those in Figs.~\ref{ARPES}f and ~\ref{ARPES}g, respectively. In the data of the 5-layer compound (Fig.~\ref{ARPES}f), the separation of two pockets (IP$_0$ and IP$_1$) is not very clear likely because of the limited momentum resolution in synchrotron-ARPES. These pockets are so close to each other that they mutually overlap in spectra, consequently showing a relatively broad feature. In contrast, two Fermi pockets are clearly distinguished in the ARPES mapping for the 6-layer compound (Fig.~\ref{ARPES}g) because of a more significant difference between the sizes of the two pockets. These data by synchrotron ARPES are consistent with those obtained by a laser-ARPES showing that the small Fermi pocket (IP$_0$) of the 6-layer compound gets half that in area of the 5-layer compound, whereas the large Fermi pocket (IP$_1$) and Fermi arc (OP) are unchanged between the two compounds.\\

{\noindent \fontsize{13pt}{13pt}\selectfont \textbf {Note 3: Model calculations demonstrating that CuO$_2$ planes independently form different Fermi surfaces (Fig. S3)}}
\vskip.5\baselineskip
%\section {Note 3: M\MakeLowercase{odel calculations demonstrating that} C\MakeLowercase{u}O$_2$ \MakeLowercase{planes independently form different} F\MakeLowercase{ermi surfaces} (F\MakeLowercase{ig.} \ref{calc})}
In the main text, we estimated the carrier concentrations ($p$s) of the innermost CuO$_2$ planes (IP$_0$) and the 2nd inner planes (IP$_1$) each from the area of the small and large Fermi pocket. This is based on the assumption that the mixing of wave functions among layers is negligible, so the CuO$_2$ planes independently form different Fermi surfaces. Here, we confirm that this assumption is indeed valid, by performing model calculations to examine the influence of inter-layer hopping (V), which induces the mixing of wave functions among layers, on the Fermi surfaces.

\begin{figure*}[h]
\includegraphics[clip,width=16.5cm]{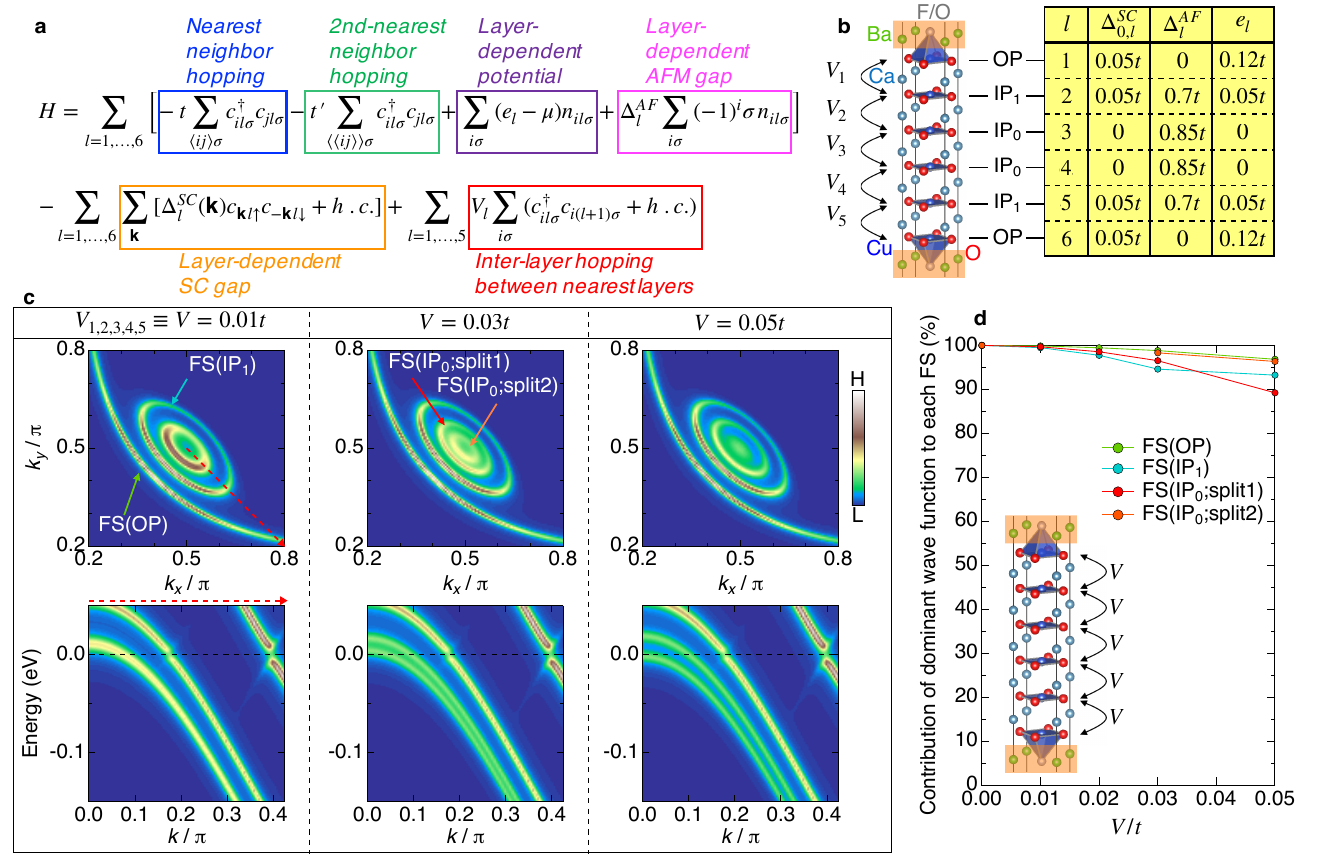}
\renewcommand{\baselinestretch}{1}
\caption{\textbf{Model calculations demonstrating that CuO$_2$ planes independently form different Fermi surfaces.} \textbf{a,} Model Hamiltonian we used to reproduce the overall band structure of the 6-layered cuprate. The physical meaning of each term is described nearby the equation. 
 \textbf{b,}  The crystal structure of 6-layer compound and physical parameters for each layer $l$ which reproduce the spectral feature obtained by ARPES. Here, we set $t$ = 0.22 eV and $t'$ = -0.43t.  \textbf{c,} Variation of Fermi surfaces and band dispersions with increasing interlayer hopping parameter $V_l$. For simplicity, we set all values of $V_l$ to be the same $V_{1,2,3,4,5}$ $ \equiv$ $V$.
The small Fermi pocket mainly formed by the double innermost layers (IP$_0$) is most sensitive to a finite value of $V$. While it is not clear at \textit{V} = 0.01$t$ (the left panels), a band splitting is clearly seen at \textit{V} = 0.03$t$ (the middle panels), and it gets more pronounced at higher values as \textit{V} = 0.05$t$ (the right panels). 
 Arrows in \textbf{d} point to Fermi surfaces with naming such as FS(IP$_1$) for the Fermi surface mainly formed by the wave function of IP$_1$. 
\textbf{d,} Contribution percentage of the wave function from the dominant layer to each Fermi surface. Even at $V$ = 0.03$t$, which yields a band splitting so large as not to be observed, the mixing among layers is small, as more than 90~$\%$ of the contribution comes from the main layer. The contribution at $V$ = 0.02$t$ or less, closer to reality, reaches nearly 100~$\%$, indicating that each layer forms the Fermi surface independently from other layers in real materials.}
\label{calc}
\end{figure*}

Figure \ref{calc}a shows the Hamiltonian used for the model calculation, which expresses the band structure of a system with six CuO$_2$ layers per unit cell indicated by $l=1-6$. The physical meaning for each term is described beside the equation. The overall spectral structure obtained by ARPES is well reproduced with $t=0.22$ eV and $t'=-0.43t$ and by setting the layer-dependent values of potential $e _l$, superconducting gap  {$\Delta_l^{SC}(k)=\Delta_{0,l}^{SC}$$({\rm cos~}k_x-{\rm cos}~k_y)/2$}, and antiferromagnetic gap $\Delta _l^{AF}$, as listed in Fig. \ref{calc}b. The mixing of wave functions among layers can be induced by the inclusion of inter-layer hopping ($V_l$; $l$=1-5). However, we find that it is very small because otherwise extra band splitting would occur, which disagrees with our experimental observation, as detailed below.

%However, this value must be very small because otherwise extra band splitting would occur and not match the observed reality.
%{$\Delta_l^{SC}(k)=\frac{\Delta_{0,l}^{SC}}{2}$$({\rm cos~}k_x-{\rm cos}~k_y)$}

For simplicity, we set all $V_l$ to have the same value $V_{1,2,3,4,5}$ $\equiv$ $V$. In Fig. \ref{calc}c, we plot the calculated Fermi surfaces and energy dispersions along the antiferromagnetic zone boundary (AFZB) for three cases with different $V$s of $V=0.01t$ (the left panels),  $V=0.03t$ (the middle panels), and  $V=0.05t$ (the right panels). The six-layer compound has doubled innermost layers (IP$_0$) and the electronic states of these adjacent layers with the same potentials ($e_3=e_4$) should be most sensitive to the inclusion of $V$. The bilayer splitting for IP$_0$, therefore, is expected to appear even with a small value of $V$. Indeed, we found that only $V=0.03t$ is enough to generate a clear splitting due to the bilayer. However, neither ARPES nor quantum oscillation measurements show such a splitting, thus we can conclude that $V$ should be much less than $0.03t$ in the real materials. We also note that, if the splitting exists even for a small amount, we would see the associated broadening in the ARPES spectra and the FFT spectra of quantum oscillation for the small pocket of IP$_0$, compared with those for the large pocket of IP$_1$. Such an indication, however, is not seen in both of the experiments; hence, $V$ should be very small as $0.01t$ at most.  

In Fig. \ref{calc}d, we calculated the contribution percentage of the wave function in the dominant layer to each Fermi surface:  FS(OP), FS(IP$_1$), FS(IP$_0$; split1), and FS(IP$_0$; split2), which are named in the top panels of Fig. \ref{calc}c. 
For example, the light blue circles in Fig. \ref{calc}d represent the percentage of wave function distributed by the 2nd inner plane (IP$_1$) to form FS(IP$_1$); likewise, the red circles represent the percentage by the innermost plane (IP$_0$) to form FS(IP$_0$; split1). As a whole, high values are obtained for either case, meaning that the mixing of wave functions among planes is very small. Notably, even at $V=0.03t$, which generates an unrealistically large bilayer splitting (see the right panels of Fig. \ref{calc}c), more than 90~$\%$ of the spectral weight is contributed from the main CuO$_2$ layer. 
Most importantly, the percentage increases up to 97~$\%$ at $V=0.02t$, which is the expected upper limit of the V value according to our data. Hence, we can conclude that, in a real material, each CuO$_2$ layer independently forms different Fermi surfaces corresponding to each doping level. 
This justifies the means employed in the main paper to estimate the carrier concentration ($p$)
for the inner CuO$_2$ planes (IP$_1$ and IP$_0$) independently from the areas of small and large Fermi pockets, respectively.\\

{\noindent \fontsize{13pt}{13pt}\selectfont \textbf {Note 4: Comparison of superconducting gaps among three Fermi surfaces for OP, IP1, and IP0 (Fig. S4)}}
\vskip.5\baselineskip
%\section {Note 4: C\MakeLowercase{omparison of superconducting gaps among three} F\MakeLowercase{ermi surfaces for} OP, IP$_1$, \MakeLowercase{and} IP$_0$ (F\MakeLowercase{ig.} \ref{gap})}
In the main paper, we present the superconducting gap only for the 2nd inner planes (IP$_1$) with a large Fermi pocket, and also the data points are only for a few $k_F$ points. To fully understand the superconducting properties in the 6-layered cuprates, we examine here all the results of the superconducting gaps including those for the outer and innermost planes (OP and IP$_0$), which form the Fermi arc and small Fermi pocket, respectively.

\begin{figure*}[h]
\includegraphics[clip,width=17cm]{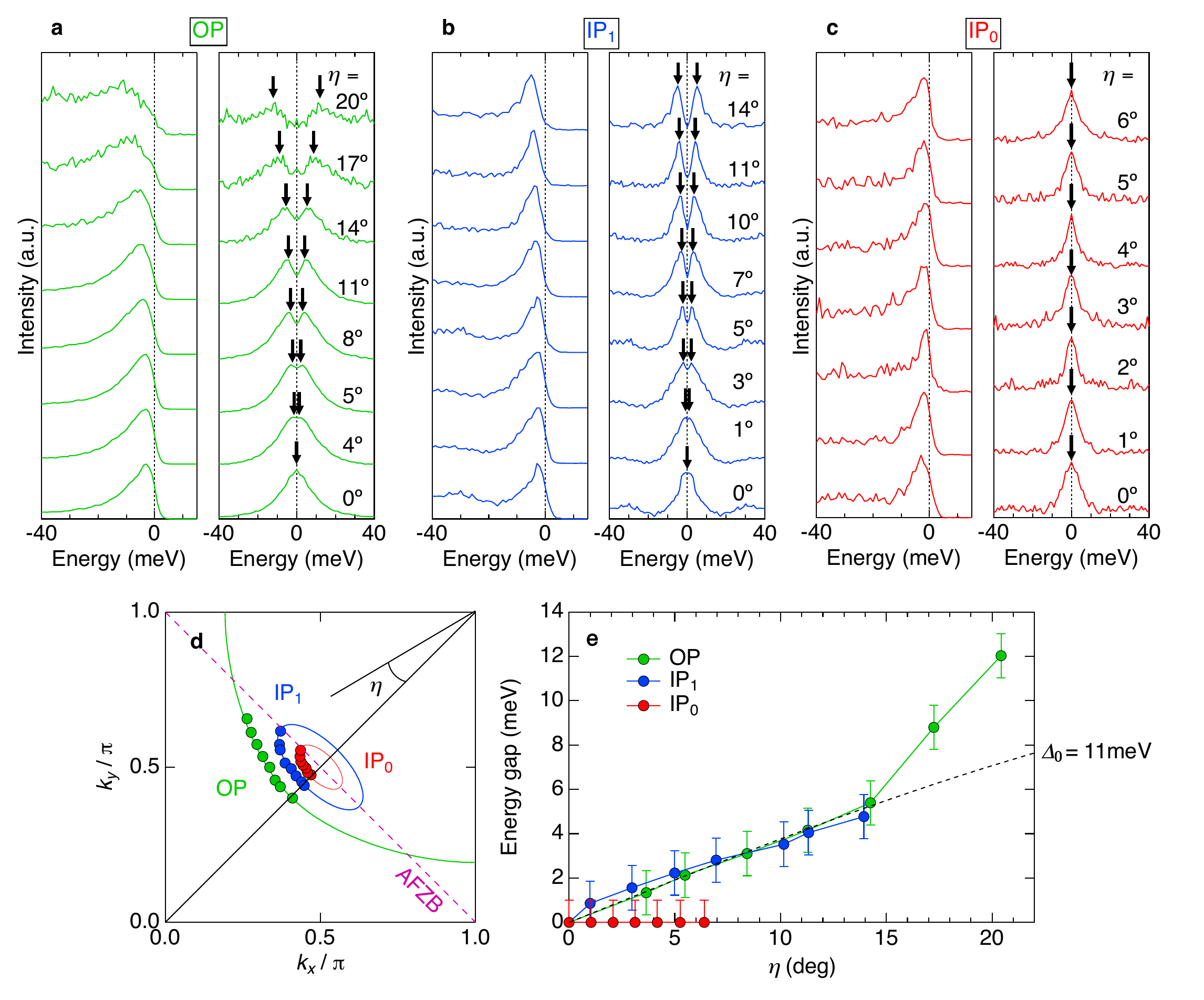}
\renewcommand{\baselinestretch}{1}
\caption{\textbf{Comparison of the superconducting gaps among three Fermi surfaces for OP, IP$_1$, and IP$_0$.} 
\textbf{a-c,} Energy distribution curves (EDCs) and those symmetrized about the Fermi level,  
measured around the gap node for the Fermi arc (OP) and pockets (IP$_1$ and IP$_0$).
The arrows point to the spectral peaks. The \textit{k}$_F$ points measured for each Fermi surface are marked by colored circles in \textbf{d}. 
\textbf{d,} Fermi surfaces (solid lines) determined by the tight-binding fitting to ARPES data and \textit{k}$_F$ points where the EDCs of \textbf{a-c} were measured. 
\textbf{e,} Angle dependence of the energy gap for three Fermi surfaces (OP, IP$_1$, and IP$_0$). The angle $\eta$ is defined in \textbf{d}. $\Delta_0=11$meV is estimated by extrapolating the gaps near the node up to the antinode.
The gap is absent for IP$_0$ all around the Fermi pocket, as confirmed in the symmetrized EDCs (the right panel of \textbf{c}), which all have a single peak at the Fermi level. Error bars represent standard deviations of the spectral peak positions.
} 
\label{gap}
\end{figure*}

The left panels of Figs. \ref{gap}a, \ref{gap}b, and \ref{gap}c plot energy distribution curves (EDCs) measured at $k_F$s along the Fermi surface for OP (Fermi arc), IP$_1$ (large Fermi pocket), and  IP$_0$ (small Fermi pocket) around the gap node (green, blue, and red circles in Fig. \ref{gap}d), respectively. In the right panels of Figs. \ref{gap}a, \ref{gap}b, and \ref{gap}c, we symmetrize these spectra about the Fermi level to eliminate the effect of the Fermi cut-off and visualize a gap opening (or gap closing). A $d$-wave-like gap opens in OP (Fig. \ref{gap}a) and IP$_1$ (Fig. \ref{gap}b), as traced by arrows pointing to the peak positions of spectra: the spectral gap is zero at $\eta$=0$^\circ$ ($\eta$ is defined in Fig. \ref{gap}d), and it opens off the gap node, increasing with larger $\eta$s toward the antinode. In contrast, we observe no gap all around the Fermi surface (or Fermi pocket) for IP$_0$ (see Fig. \ref{gap}c) within the experimental energy resolution.

The superconducting gaps estimated against the $\eta$ angle are summarized in Fig. \ref{gap}e. We found mainly two notable features in the data: (1) the superconducting gaps are comparable between the Fermi arc for OP and the Fermi pocket for IP$_1$ ($\Delta_0=11$meV, estimated by extrapolating the gaps near the node up to the antinode) and (2) the superconducting gap is absent in the Fermi pocket for IP$_0$. IP$_1$ is spatially more distant from the dopant layers than OP in the crystal structure, thus it is less doped and expected to have smaller superconducting gaps than those of OP. Finding (1), that it is not the case in a real material, suggests that the electron pairing gets more stabilized in the Fermi pocket, which is formed by clean CuO$_2$ planes. Another possible reason for it is that the pocket can avoid competition with other ordered states (pseudogap and charge-density-wave states). Although these competing orders tend to develop near ($\pi$,0), thus they cannot develop since low-lying electronic states required to form those are lacking in IP$_1$ which form the Fermi surface (or Fermi pocket) only around the node. 

On the other hand, Finding (2) has two implications as follows. First, the two pockets are independently formed by IP$_0$ and IP$_1$ since otherwise, the mixing of layers would produce superconducting gaps of similar magnitudes for both the pockets, unlike in our observation. Secondly, the electronic state of IP$_0$ with less doping than IP$_1$ is situated outside of the superconducting dome in the phase diagram.\\

\begin{figure*}[h]
\includegraphics[clip,width=16cm]{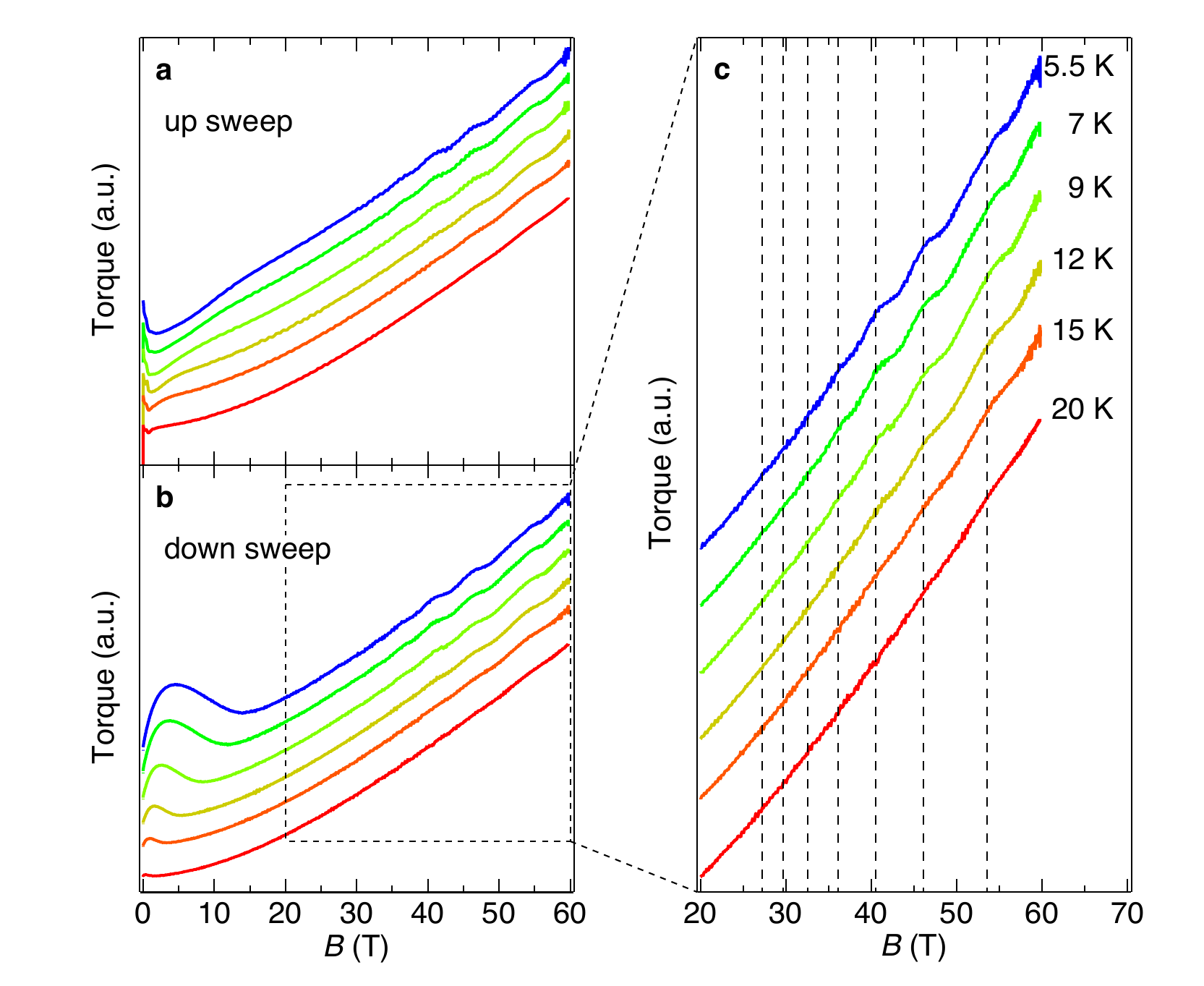}
\renewcommand{\baselinestretch}{1}
\caption{{\bf Raw data of de Haas-van Alphen effect.}
\textbf{a,b,} Magnetic torque signals (de Haas-van Alphen effect: dHvA) at several temperatures detected during up sweep and down sweep of a pulsed magnetic field up to 60 T, respectively. An offset is used in these plots to clearly display the behavior of each curve. 
The angle between the applied magnetic field and the crystallographic $c$-axis was set to be 2 degrees. The broad hump and dip seen at low fields in the up sweep and down sweep measurements, respectively, 
are attributable to the irreversibility field. 
\textbf{c,}  The zoom of the dashed area in \textbf{b}. The peaks of oscillations are indicated by dashed lines.}
\label{raw_dHvA}
\end{figure*}

{\noindent \fontsize{13pt}{13pt}\selectfont \textbf {Note 5: Raw data of Haas-van Alphen effect (Fig. S5)}}
\vskip.5\baselineskip
%\section {Note 5:  R\MakeLowercase{aw data of} H\MakeLowercase{aas-van} A\MakeLowercase{lphen effect} (F\MakeLowercase{ig.} \ref{raw_dHvA})}
In the main paper, we show quantum oscillations of magnetic torque signals (de Haas-van Alphen effect: dHvA) after background subtraction. The background was obtained by fitting a quadratic function to each curve of the raw data between 26 and 60 T. Here we exhibit the raw magnetic toque signals before background subtraction measured during the up sweep (Fig. \ref{raw_dHvA}a) and the down sweep (Fig. \ref{raw_dHvA}b) of a pulsed magnetic field. 
The behaviors of the paired two curves for each temperature are different in the low range of magnetic field (B) less than the lower critical field ($H_{c1}$) with a dip and hump attributed to the irreducibility field for the up sweep and down sweep measurements, respectively. At magnetic fields higher than $H_{c1}$, the two curves match each other with showing oscillations.
Although the oscillation amplitudes against $B$ are not very high, it is still clearly visible above $\sim$30T (see zoomed data shown in Fig. \ref{raw_dHvA}c), which is sufficient for the analysis to extract the intrinsic frequencies.\\

{\noindent \fontsize{13pt}{13pt}\selectfont \textbf {Note 6: Effective mass and Dingle temperature determined by de Haas-van Alphen effect (Fig. S6)}}
\vskip.5\baselineskip
%\section {Note 6: E\MakeLowercase{ffective mass and} D\MakeLowercase{ingle temperature determined by de} H\MakeLowercase{aas-van} A\MakeLowercase{lphen effect} (F\MakeLowercase{ig.} \ref{Dingle})}
From the behavior of quantum oscillations, we can extract two physical quantities: effective mass ($m^*$) and Dingle temperature ($T_D$). Here we estimate these values from the data of de Haas-van Alphen effect (dHvA) exhibiting clear oscillations (the main Fig. 1c). In particular, we find here that the inner planes of the 6-layer compound are indeed very clean. This agrees that the ARPES spectra show quasiparticle sharp peaks even in the lightly doped region, where the screening effect gets so weak that the coherence of conducting electrons could be deprived even by the slightest disorder.

Here, we focus only on the large Fermi pocket formed by the 2nd inner CuO$_2$ plane (IP$_1$) for the estimation of $m^*$ and $T_D$; note that the maximum magnetic field (60 T) we applied was not high enough to estimate $T_D$ of the small Fermi pocket for the innermost CuO$_2$ plane (IP$_0$), since this tiny pocket contributes to a lower frequency component in our data of the dHvA effect and it displays only a limited number of oscillations within the magnetic range up to 60 T, not allowing a reliable estimation of $T_D$. 

\begin{figure*}[h]
\includegraphics[clip,width=16cm]{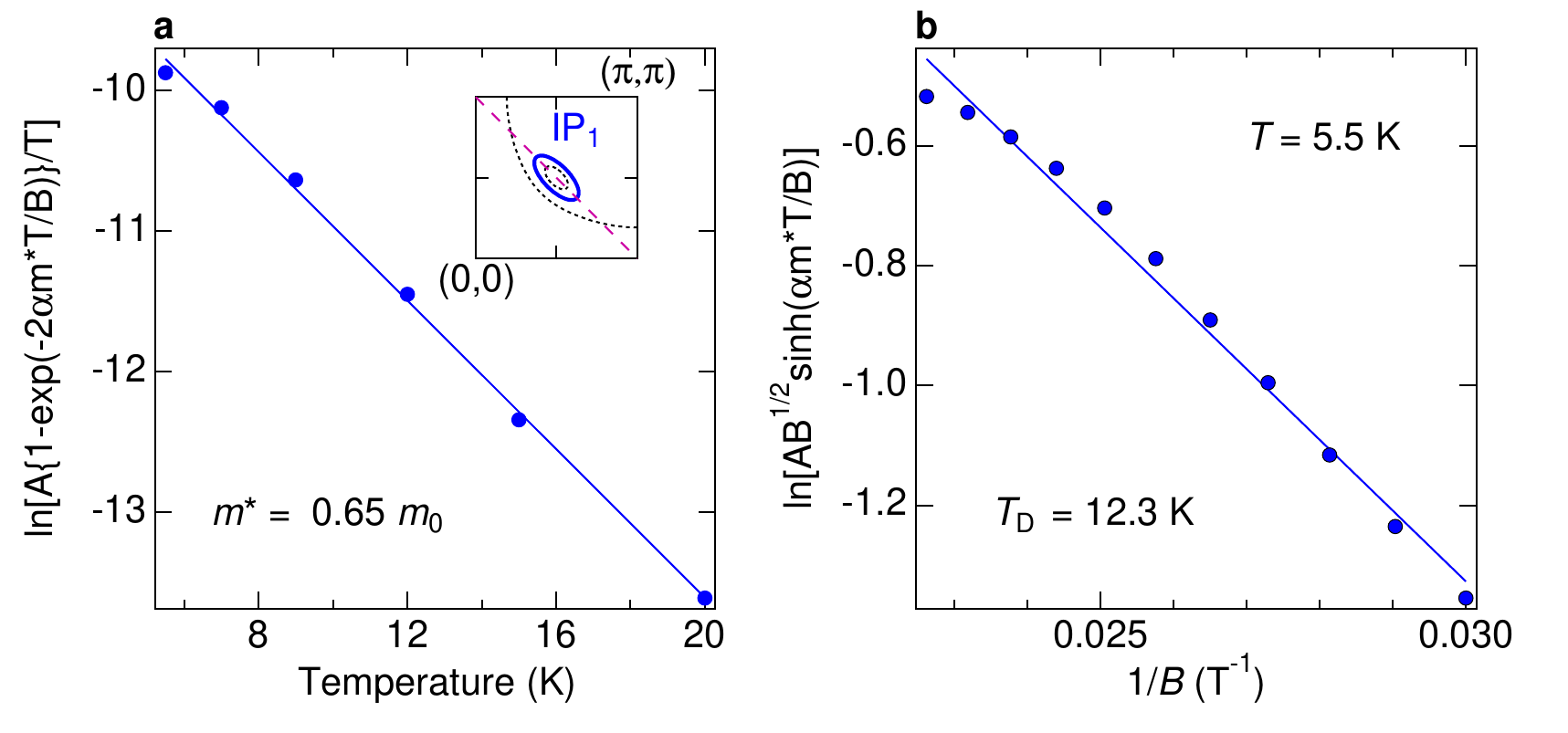}
\renewcommand{\baselinestretch}{1}
\caption{\textbf{Effective mass and Dingle temperature for the large Fermi pocket (IP$_1$) estimated from quantum oscillation measurements.} \textbf{a,} Temperature dependence of the quantum oscillation amplitude. The mass plot with the effective mass ($m*$) of 0.65 m$_0$ best fits with the standard Lifshitz-Kosevich formula. \textbf{b,} Dingle plot as a function of 1/\textit{B}. Dingle temperature (\textit{T}$_D$) of 12.3 K is obtained by fitting the data.}
\label{Dingle}
\end{figure*}

Figure \ref{Dingle}a plots the fast Fourier transformation (FFT) amplitude of the dHvA oscillation against temperature. The fitting of the data to the standard Lifshitz-Kosevich formula yields an effective mass ($m^*$) of 0.65 $m_0$ \cite{Akiba} ($m_0$: the free electron mass). This is consistent with the effective mass of Fermi pocket observed by ARPES. From the slope of Dingle plot against 1/$B$ (Fig. \ref{Dingle}b), one can estimate 
the value of $T_D$, which is proportional to the scattering rate of conducting electrons. 
By the fitting to the data at $T=$5.5 K, we obtained $T_D$ of 12.3 K. Notably, this value is comparable to that for IP$_1$ of the 5-layer compound (\textit{T}$_D$ = 11.8K) \cite{Kunisada2020}, or between those of YBa$_2$Cu$_3$O$_y$ (\textit{T}$_D$ = 6.2K) and HgBa$_2$CuO$_{4+\delta}$ (\textit{T}$_D$ = 18K) \cite{Chan2016}, which are thought to have very clean CuO$_2$ planes. Note that the carrier concentration of IP$_1$ directly estimated from the Fermi pocket area is small, only to be 4.3~$\%$, which is less than half those ($\sim$10~$\%$) of Y123 and Hg1201 samples used for the quantum oscillation measurements. Hence, the \textit{T}$_D$ of IP$_1$ is rather small, considering that it is obtained in such a low carrier concentration with a poor screening effect. The $T_D$ value we obtained here, therefore, indicates that the inner CuO$_2$ layers of the 6-layer compound are very clean without disorder.\\

\begin{figure*}[h]
\includegraphics[clip,width=10cm]{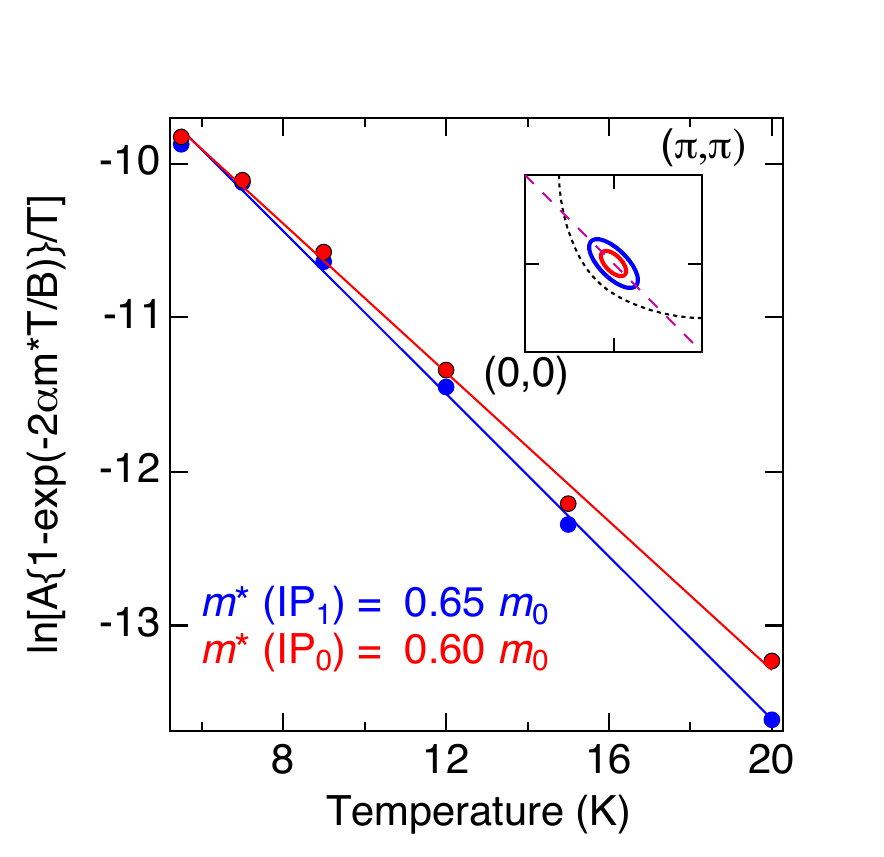}
\renewcommand{\baselinestretch}{1}
\caption{\textbf{Almost the same effective mass for the small Fermi pocket (IP$_0$) and the large Fermi pocket (IP$_1$) estimated from quantum oscillation measurements.} The mass plot fitted with the standard Lifshitz-Kosevich formula estimates the effective mass ($m*$) to be 0.60 m$_0$ and 0.65 m$_0$ 
for the small Fermi pocket (IP$_0$) and the large Fermi pocket (IP$_1$), respectively. Here, the data for IP$_1$ is duplicated from Fig.~\ref{Dingle}a.}
\label{mass_comparison}
\end{figure*}

{\noindent \fontsize{13pt}{13pt}\selectfont \textbf {Note 7: Doping independent effective mass in the lightly doped CuO$_2$ planes (Fig. S7)}}
\vskip.5\baselineskip
%\section {Note 7: D\MakeLowercase{oping independent effective mass in the lightly doped} C\MakeLowercase{u}O$_2$ \MakeLowercase{planes} (F\MakeLowercase{ig.} \ref{mass_comparison})}
The doping dependence of correlation effects is important to understand the metal to Mott insulator transition in cuprates. 
While the K deposition (controlling carrier concentrations) does not change the peak width of spectra (lifetime of quasiparticles), it inevitably degrades the sample surface, reducing the peak-to-background ratio with deposition time. This may mislead one when understanding the correlation effects with different doping levels, such as the behavior of quasiparticle residue. The best way for this study is to use samples with different carrier concentrations and compare their ARPES data taken all from freshly cleaved surfaces. At present, however, we have not succeeded in preparing such crystals. Instead, we compare the effective masses ($m^*$) of band structures
for the innermost layer (IP$_0$) and the 2nd inner layer (IP$_1$), which have different carrier concentrations $p$ ($\sim$1~$\%$ and $\sim$4~$\%$, respectively).  The effective masse is sensitive to electron correlations, so this comparison 
tells us how the electron correlation effect at different $p$s leads to the metal to Mott insulator transition at the half-filling. The $m^*$s are determined by the mass plots of the quantum oscillation data for two Fermi pockets [Fig.~\ref{mass_comparison}(a,b)]. Interestingly, we found that they are almost the same ($\sim$0.6 m$_0$). This indicates that a pronounced band narrowing does not take place with reducing $p$ toward the half-filling. The transition to a Mott insulator most likely occurs by completely removing hole carriers from the lower Hubbard band until the perfect half-filling, rather than by controlling the band width.\\

\begin{figure*}[h]
\includegraphics[clip,width=6.5cm]{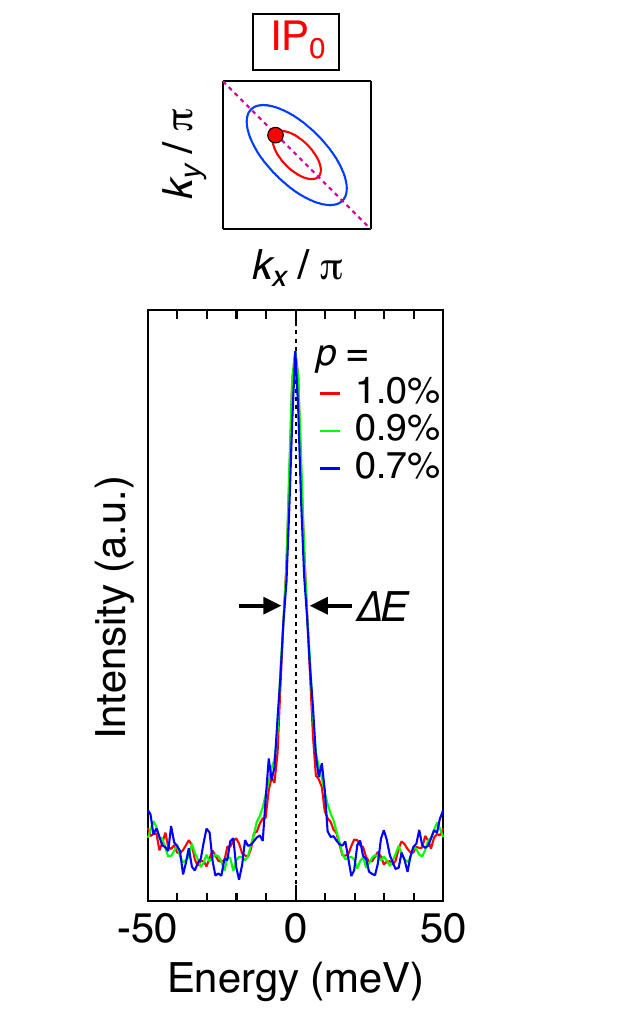}
\renewcommand{\baselinestretch}{1}
\caption{\textbf{No change in the peak width of the nodal spectra for different carrier concentrations.} The spectral peaks (symmetrized EDCs) for different carrier concentrations ($p$ = 1.0~$\%$,  0.9~$\%$, and 0.7~$\%$) controlled by K deposition are compared for the innermost plane (IP$_0$). These spectral widths are almost the same, and thus scattering rate (or lifetime) of quasiparticles does not change. This validates that long-lived, well-defined quasiparticles are established even at an extremely lightly doped state $p$ = 0.7~$\%$, in the close vicinity of a half-filled Mott state.}
\label{QP1}
\end{figure*}

{\noindent \fontsize{13pt}{13pt}\selectfont \textbf {Note 8: Long-lived, well-defined quasiparticle at an extremely low carrier concentration of $p$ = 0.7 $\%$ (Fig. S8)}}
\vskip.5\baselineskip
%\section {Note 8:  L\MakeLowercase{ong-lived, well-defined quasiparticle at an extremely low carrier concentration of p~=~0.7~$\%$} (F\MakeLowercase{ig.} \ref{QP1})}
In the main text, we argue that the quasiparticles are well-defined even at doping levels much less than 1~$\%$. Here we demonstrate that the peak width (lifetime of quasiparticles) does not vary by reducing $p$ from 1.0~$\%$ to 0.7~$\%$. In Fig. \ref{QP1}, we directly compare the spectral peaks of symmetrized EDCs for small Fermi pocket (IP$_0$) with three different doping levels controlled by K deposition: $p$ = 1.0~$\%$, 0.9~$\%$, and 0.7~$\%$. These peak shapes and widths are almost identical; that is, the scattering rates (or lifetimes) of quasiparticles are unchanged by reducing $p$ down to 0.7~$\%$, and the quasiparticles are well-defined with a long lifetime even extremely close to half-filled Mott state.\\

{\noindent \fontsize{13pt}{13pt}\selectfont \textbf {Note 9: Closing of an energy gap around $p$ = 4.0 $\%$, the edge of superconducting dome (Fig. S9)}}
\vskip.5\baselineskip
%\section {Note 9: C\MakeLowercase{losing of an energy gap around p = 4.0~$\%$, the edge of superconducting dome} (F\MakeLowercase{ig.} \ref{QP2})}
Although the energy resolution of our experiments is very high ($\sim$1.4 meV) owing to a laser ARPES, it is still true and never be avoided that experiments have finite resolutions. Accepting such a reality, still, our conclusion that the energy gap is closed or negligible at doping levels less than $p$ = 4.0~$\%$ (outside of the superconducting dome) is reasonably led by ``theoretical simulation" and ``close examinations of ARPES spectra" described below.\\

\textbf{Theoretical simulation:}
In our previous publication~\cite{Kunisada2020}, the superconducting proximity effect was confirmed to be negligible in multilayer cuprates by simulation with a model Hamiltonian reproducing the band structure of a 5-layer compound obtained by ARPES. In the current work (supplemental Fig. S3), we further confirmed by simulation reproducing ARPES data that the interlayer mixing of the wave function is negligible: it was found that more than 97~$\%$ of the wave function for each band comes dominantly from one of the multiple CuO$_2$ layers. All these simulations based on ARPES data strongly indicate that the superconducting proximity effect in multilayer cuprates is negligible. Indeed, it is never possible to prove that the proximity effect is absolutely zero in any experiments that are inevitably limited by finite resolutions. However, it would be reasonable enough to mention that it should be negligible based on the simulation and the data compatible with it.\\

\textbf{Close examinations of ARPES spectra:}
To justify our conclusion further, here we closely examine the peak width of symmetrized EDC usually used for evaluating the gap opening. We point out that the spectral peak in the inner layer of 6-layer cuprates is even sharper than that of the optimally doped Bi2212 measured by a laser ARPES with the same experimental condition. Such sharp spectra are very sensitive to a gap opening or closing: we can tell that a gap is opened when the peak width gets broadened, or a gap is closed when the peak width is sharpened, as a function of doping and Fermi angle.

\begin{figure*}[h]
\includegraphics[clip,width=16cm]{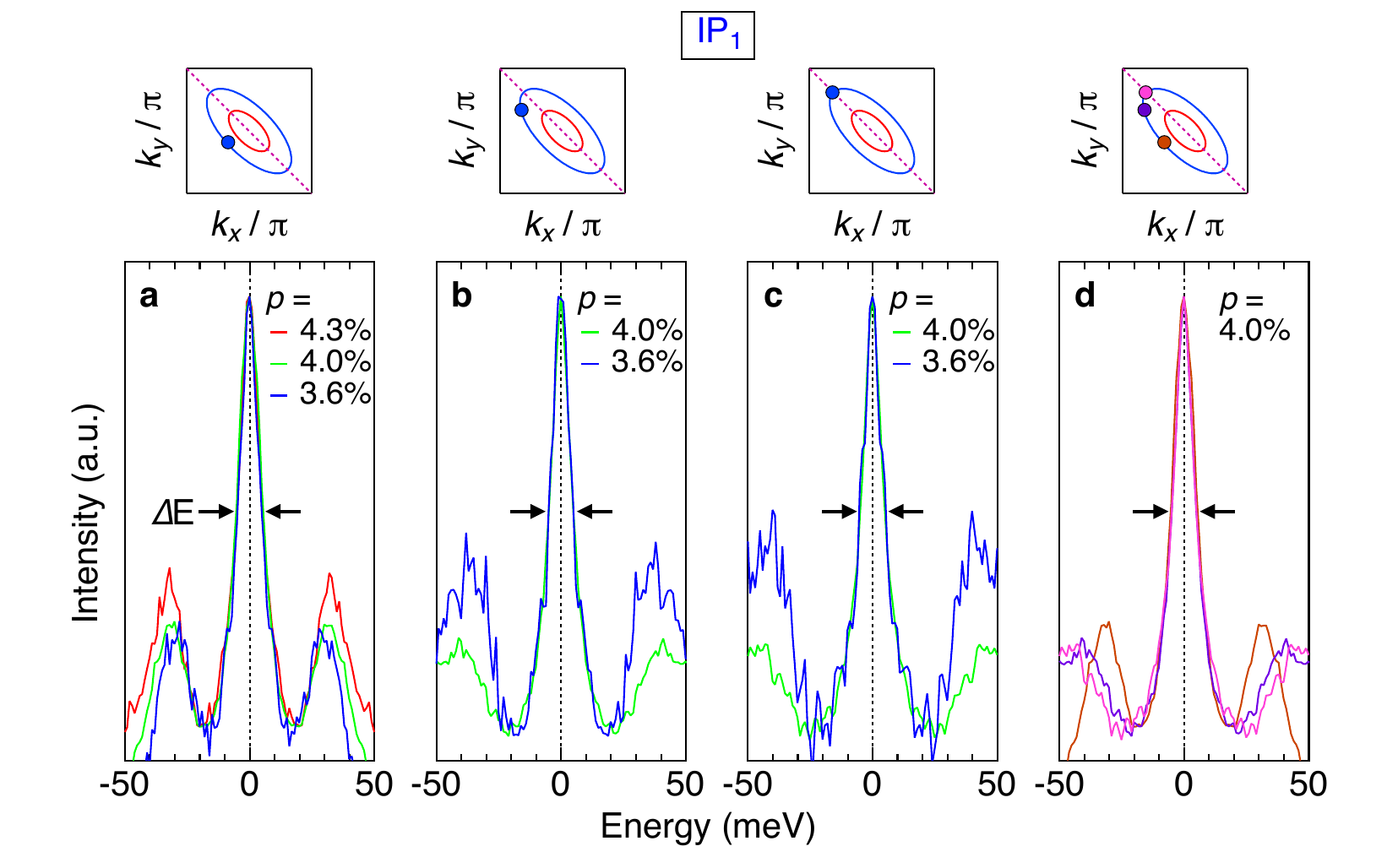}
\renewcommand{\baselinestretch}{1}
\caption{\textbf{Validating no superconducting gap at carrier concentrations less than 4.0~$\%$} \textbf{a,} The spectral peaks (symmetrized EDCs at the nodal $k_{\rm F}$) for different carrier concentrations ($p$ = 4.3~$\%$,  4.0~$\%$, and 3.6~$\%$) controlled by K deposition are compared for the 2nd inner plane (IP$_1$). Here, the $p$ for the pristine surface is 4.3~$\%$.  Almost the same peak widths are confirmed. \textbf{b,} The spectral peaks at $p$ = 4.0~$\%$ and 3.6~$\%$ for a $k_{\rm F}$ between the node and the tip of the Fermi pocket (circle in the inset). Almost the same peak widths are confirmed. \textbf{c,} The same data as \textbf{b}, but obtained at the tip of the Fermi pocket (circle in the inset). Almost the same peak widths are confirmed. \textbf{d,} Spectral peaks at three $k_{\rm F}$ points (circles in the inset). Almost the same peak widths are confirmed. These results (no variation of peak width with doping and Fermi angle) justify that there is no superconducting gap at doping levels less than $p$~=~4.0~$\%$.}
\label{QP2}
\end{figure*}

First of all, let’s check the spectral width at the nodal $k_{\rm F}$ point, where it is accepted that there is no gap. Figure~\ref{QP2}a compares the spectral peak shape at the nodal point of the larger Fermi pocket (IP$_1$) with three different hole-carrier concentrations ($p$ = 4.3~$\%$, 4.0~$\%$, and 3.6~$\%$) controlled by K deposition. $p$ = 4.3~$\%$ is for the pristine surface which opens a superconducting gap at the $k_{\rm F}$ points off the node (see the main Fig. 4e and 4f). $p$ = 4.0~$\%$ and 3.6~$\%$ are located outside the superconducting dome, according to our evaluation. The spectral peak widths are found to be almost the same regardless of the doping levels. These spectra at the node, therefore, can be used as comparison references to judge whether or not a gap is open at $k_{\rm F}$ points off the node.

Now we turn to the spectra at $k_{\rm F}$’s off the nodal direction. Figures \ref{QP2}b and \ref{QP2}c compare the spectral peaks at $p$ = 4.0~$\%$ and 3.6~$\%$ (outside the superconducting dome) for two different $k_{\rm F}$ points marked on the Fermi pocket (IP$_1$) in each upper panel: one is located between the node and the tip of the oval-shaped Fermi pocket (Fig. \ref{QP2}b), and the other is for the tip (Fig. \ref{QP2}c). Similarly to the case of the nodal direction (Fig. \ref{QP2}a), we find that the spectral width is almost the same for $p$ = 4.0~$\%$ and 3.6~$\%$ at both $k_{\rm F}$ points.  In Fig. \ref{QP2}d, we further compare the peak widths for different $k_{\rm F}$’s using the spectra at $p$ = 4.0~$\%$ and find that these are all the same. These results reasonably lead us to conclude that there is no energy gap at $p$ = 4.0~$\%$ and 3.6~$\%$ within the energy resolution. Our data rather indicate that the scattering rate, which can be estimated from the spectral width, is isotropic. 
Since a clear gap is observed at $p$ = 4.3~$\%$, in contrast to $p$ = 4.0~$\%$ and 3.6~$\%$, the edge of the superconducting dome is decided as $\sim$4.0~$\%$, as argued in the main text.\\

\begin{figure*}[h]
\includegraphics[clip,width=16cm]{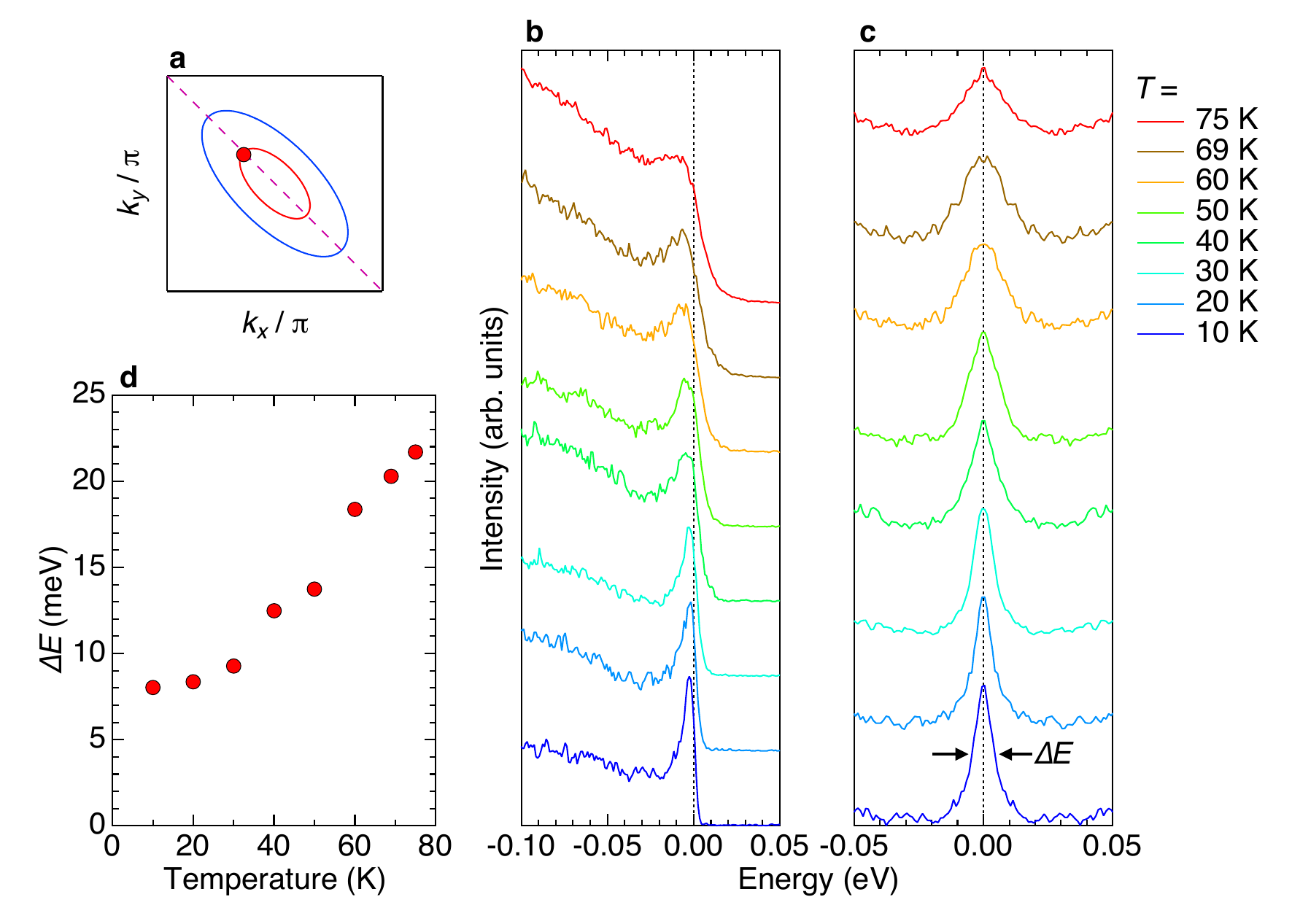}
\renewcommand{\baselinestretch}{1}
\caption{\textbf{Persisting quasiparticle picture above $T_c$ at an extremely low carrier concentration of $1\%$.} \textbf{a,} Schematic  Fermi pockets and the measured $k_{\rm F}$ point (marked by circle). \textbf{b,c} The temperature evolution of EDCs and those symmetrized, respectively, measured from 10K to 75K above $T_c$~=~69~K. 
\textbf{d,} The temperature evolution of the spectral peak width (arrows in \textbf{c}). No particular anomaly is observed over the temperatures through $T_c$ other than thermal broadening, confirming that quasiparticles are not products of bulk superconductivity.}
\label{Tdep}
\end{figure*}

{\noindent \fontsize{13pt}{13pt}\selectfont \textbf {Note 10: Evidence for quasiparticles well-defined without the infludence of superconducitity (Fig. S10)}}
\vskip.5\baselineskip
%\section {Note 10:  E\MakeLowercase{vidence for quasiparticles well-defined without the infludence of superconducitity} (F\MakeLowercase{ig.} \ref{Tdep})}
Here we argue that the well-defined quasiparticle peaks we observed by ARPES for the small Fermi pocket with a carrier concentration only of $p$ = 1.0~$\%$ are not generated by superconductivity. In particular, we focus on the spectra at the tip of the Fermi pocket, where the effect of superconductivity is expected to be most pronounced if there was any. The obtained spectra (raw EDCs and the symmetrized ones) measured from 10K to 75K above $T_c$ (= 69K) are plotted in Figs. \ref{Tdep}b and \ref{Tdep}c. Although the spectral peak gets broader with increasing temperature, the quasiparticle peak persists even above $T_c$, supporting our conclusion. Note that the spectral broadening at high temperatures above $T_c$ is reasonable, considering that the carrier concentration is only 1~$\%$, which is so close to the half-filled Mott state that the electronic states 
could be significantly scattered due to strong correlation effects even near $E_F$ at finite temperature.

The measurement of the de Haas-Van Alphen effect (dHvA) is a more general way to investigate quasiparticle properties with superconductivity removed. In this technique, the superconductivity is completely eliminated by a magnetic field, so one can reveal the nature of quasiparticles at low temperatures under the condition that there is no influence of superconductivity on the electronic states. Importantly, we observed a quantum oscillation reproducing the small Fermi pockets observed by ARPES. This is the strongest evidence that well-defined quasiparticles are established with nothing to do with superconductivity in the innermost layer of 6-layer cuprates.\\

\clearpage

\clearpage

\clearpage

\clearpage

\clearpage

\clearpage

\clearpage

\clearpage

\clearpage

\clearpage

\clearpage

%\bibliography{6_layer_TK}